\documentclass[a4paper,11pt]{article}
\pdfoutput=1 
\usepackage{jheppub} 

\usepackage[T1]{fontenc} 

\usepackage{bm}

\newcommand{\g}{\gamma}
\newcommand{\sig}{\sigma}

\newcommand{\open}{\sphericalangle}

\newcommand{\nn}{\nonumber}

\title{First extraction of valence transversities in a collinear framework}

\author[a,b]{Alessandro Bacchetta}
\author[c,d]{A.~Courtoy}
\author[b]{Marco Radici}

\affiliation[a]{Dipartimento di Fisica, Universit\`a di Pavia,\\  via Bassi 6, I-27100 Pavia, Italy}
\affiliation[b]{INFN Sezione di Pavia,\\ via Bassi 6, I-27100 Pavia, Italy}
\affiliation[c]{IFPA, AGO Department, Universit\'e de Li\`ege,\\ B\^at. B5, Sart Tilman B-4000 Li\`ege, Belgium}
\affiliation[d]{INFN, Laboratori Nazionali di Frascati,\\ Via E. Fermi, 40, I-00044 Frascati (Roma), Italy}

\emailAdd{alessandro.bacchetta@unipv.it}
\emailAdd{aurore.courtoy@ulg.ac.be}
\emailAdd{marco.radici@pv.infn.it}

\abstract{We present an extraction of the valence transversity parton
  distributions based on an analysis of pion-pair production in deep-inelastic
  scattering off transversely polarized targets. Recently released data
  for proton and deuteron targets at HERMES and COMPASS permit a flavor
  separation of valence transversities. 
  The present extraction is performed in
  the framework of collinear factorization, where dihadron fragmentation
  functions are involved. The latter are taken from a previous analysis of
  electron-positron annihilation measurements. }

\begin{document}
\maketitle
\flushbottom

\section{Introduction}
\label{s:intro}

The distribution of quarks and gluons inside hadrons can be described by means of
parton distribution functions (PDFs). In a parton model picture, PDFs describe
combinations of number densities of quarks and gluons in a fast-moving
hadron. The knowledge of PDFs is crucial for our understanding of QCD and for the 
interpretation of high-energy experiments involving hadrons.  At leading twist, the quark 
structure of spin-half hadrons is described by three PDFs: the unpolarized distribution 
$f_1(x)$, the longitudinal polarization (helicity) distribution $g_1(x)$, and the transverse
polarization (transversity) distribution $h_1(x)$~\cite{Ralston:1979ys,Artru:1990zv,Jaffe:1991kp, Cortes:1991ja}.  
From the phenomenological point of view, the unpolarized PDFs are well-known, 
as can be evinced by the large number of parametrizations available (see, {\it e.g.}, 
Ref.~\cite{Blumlein:2012bf} and references therein). Apart from giving us invaluable 
information about the structure of nucleons, they have a fundamental importance for the 
interpretation of measurements in any hadronic colliders, {\it e.g.}, the LHC. 
The helicity PDFs are known to some extent, see {\it e.g.}
Refs.~\cite{Leader:2010rb,deFlorian:2009vb,Hirai:2003pm,Bluemlein:2002be}. 
On the other hand, the transversity distribution is poorly known (see, {\it e.g.}, 
Refs.~\cite{D'Alesio:2012pp} and references therein). This is mainly due to the fact that 
transversity can be measured only in processes with two hadrons in the initial state~\cite{Jaffe:1997yz},
{\it e.g.} proton-proton collision, or one hadron in the initial state and at least one hadron in the
final state, {\it e.g.} semi-inclusive DIS (SIDIS). 

Combining data from HERMES~\cite{Airapetian:2004tw} and  COMPASS~\cite{Ageev:2006da} 
on polarized SIDIS with one hadron in the final state, together with data from 
Belle~\cite{Abe:2005zx} on almost back-to-back emission of two hadrons in 
$e^+ e^-$ annihilations, the transversity distribution was extracted for the first time 
by the Torino group~\cite{Anselmino:2008jk}. The main difficulty of such analysis 
lies in the factorization framework used to interpret the data, since they involve  
Transverse Momentum Dependent PDFs (TMDs). In spite of 
exceptional progress in the understanding of
TMDs~\cite{Collins:2011zzd,Aybat:2011zv,Aybat:2011ge,Echevarria:2012pw,Prokudin:2012fj,Anselmino:2012re},
we have still limited information on their evolution equations, which are needed 
when analyzing measurements at very different scales.  

In this paper, we extract the transversity distribution for the valence combination 
of up and down quarks, applying for the first time an approach based on standard 
collinear factorization. We use data on SIDIS with two hadrons detected in the final state, 
where the transversity distribution is combined with the so-called Dihadron Fragmentation 
Functions (DiFFs)~\cite{Collins:1994kq,Jaffe:1998hf,Radici:2001na}. The collinear framework 
allows us to keep under control the evolution equations of DiFFs~\cite{Ceccopieri:2007ip}. 

In Sec.~\ref{s:theory}, we summarize the theoretical framework for two-hadron SIDIS. 
The parametrization of the valence transversity and 
its error analysis is described in Sec.~\ref{s:h1param}. The results are discussed in 
Sec.~\ref{s:out}. We finally illustrate the possible applications and extensions of our 
analysis  and draw our conclusions in Sec.~\ref{s:end}.


\section{Theoretical framework for two-hadron SIDIS}
\label{s:theory}

\begin{figure}[htb]
  \centering
 \includegraphics[width=0.6\textwidth]{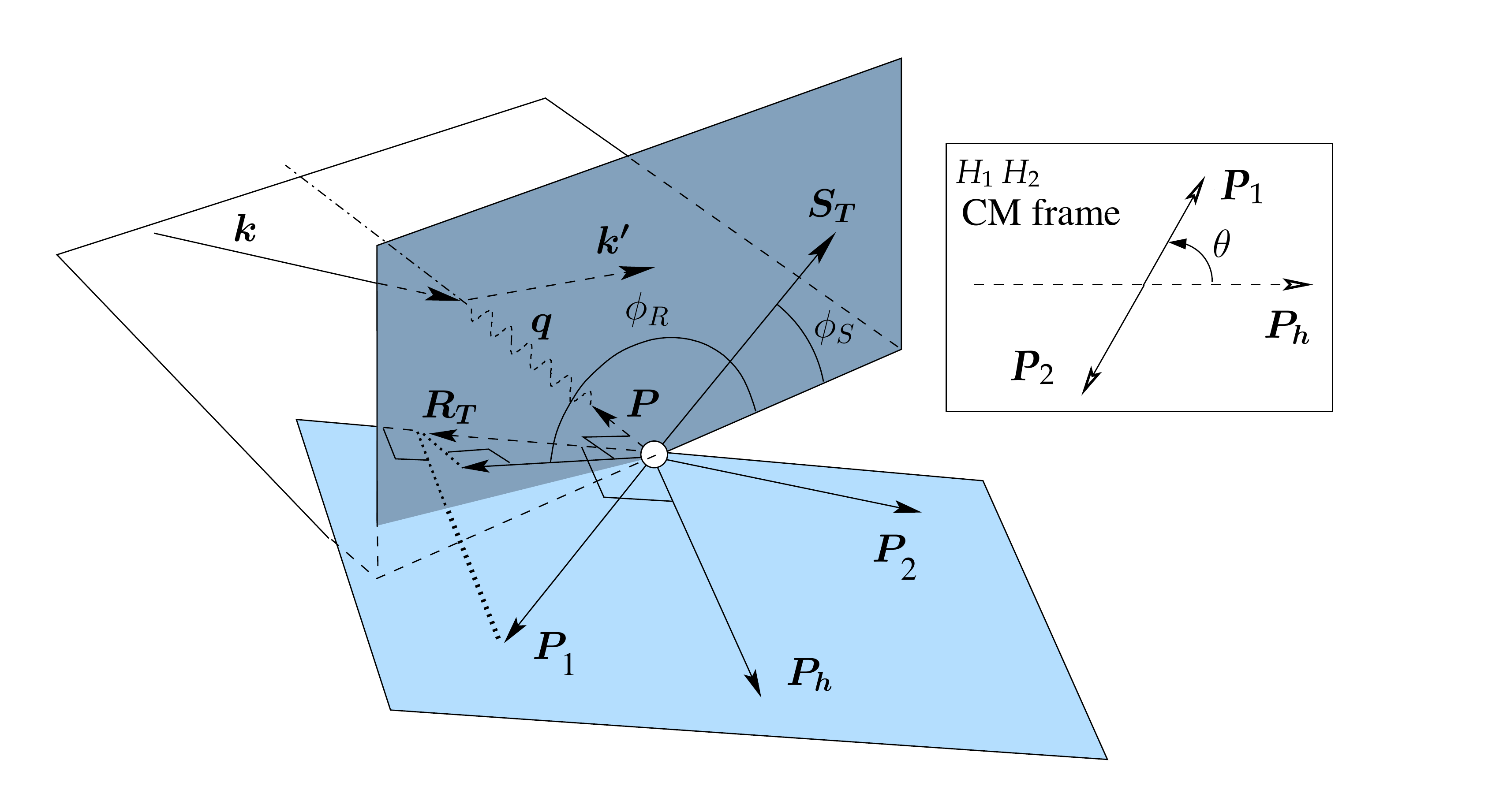}
  \caption{Kinematics of the two-hadron semi-inclusive production. The azimuthal angles 
         $\phi_{R}$ of the dihadron, and  $\phi_S$ of the component $\boldsymbol{S}_T$ of 
         the target-polarization, transverse to both the virtual-photon and target-nucleon 
         momenta $\boldsymbol{q}$ and $\boldsymbol{P}$, respectively, are 
	evaluated in the virtual-photon-nucleon center-of-momentum frame. }
  \label{f:kin}
\end{figure}

We consider the process
\begin{equation}
\ell(k) + N(P) \to \ell(k') + H_1(P_1) + H_2(P_2) + X \; ,
\label{e:2hsidis}
\end{equation}
where $\ell$ denotes the beam lepton, $N$ the nucleon target, $H_1$ and $H_2$
the produced hadrons, and where four-momenta are given in parentheses. 
We work in the one-photon exchange approximation and neglect the lepton mass.  
We denote by $M$ the mass of the nucleon and by $S$ its polarization. The final 
(unpolarized) hadrons, with mass $M_1$, $M_2$ and momenta $P_1$, $P_2$, have 
invariant mass squared $P_h^2 = M_h^2$ (which we consider much smaller than the 
hard scale $Q^2=-q^2 \geq 0$ of the SIDIS process). The $P_h = P_1 + P_2$ is the total 
momentum of the pair; we define also its relative momentum $R = (P_1-P_2)/2$. The 
momentum transferred to the nucleon target is $q = k - k'$. The kinematics of the process 
is depicted in Fig.~\ref{f:kin} (see also Ref.~\cite{Airapetian:2008sk}). 

The component 
$\bm{S}_T$ of the target polarization is transverse to both the virtual-photon 
and target momenta $\bm{q}$ and $\bm{P}$, respectively. Instead, 
$\bm{R}_T = \bm{R} - (\bm{R}\cdot \hat{\bm{P}_h})\hat{\bm{P}_h}$ is orthogonal to 
$\hat{\bm{P}_h}$ but, up to subleading-twist corrections, it can be identified 
with its projection on the plane perpendicular  to $\bm{q}$ and containing 
also $\bm{S}_T$. The azimuthal angle $\phi_{R}$ is the angle of 
$\bm{R}_T$ about the virtual-photon direction; similarly for the azimuthal 
angle $\phi_{S}$ of $\bm{S}_T$. The explicit expressions are 
\begin{align}
\phi_{R} \equiv 
&\frac{(\bm{q} \times \bm{k}) \cdot \bm{R}_T}{\vert (\bm{q} \times \bm{k}) \cdot \bm{R}_T\vert}  
\arccos {
\frac{(\bm{q} \times \bm{k}) \cdot (\bm{q}\times \bm{R}_T)}
        {\vert \bm{q} \times \bm{k}\vert \vert \bm{q}\times \bm{R}_T\vert }   }
\; , \nn \\ 
\phi_{S} \equiv
 &\frac{(\bm{q}\times \bm{k}) \cdot \bm{S}_T}{\vert (\bm{q}\times \bm{k}) \cdot \bm{S}_T\vert} 
 \arccos {
 \frac{(\bm{q} \times \bm{k}) \cdot (\bm{q} \times \bm{S}_T)}
         {\vert \bm{q}\times \bm{k}\vert \vert \bm{q} \times \bm{S}_T\vert }   }
\; . \label{e:angles}
\end{align}
We also define the polar angle $\theta$ which is the angle between the direction of the 
back-to-back emission in the center-of-mass (cm) frame of the two hadrons, and the 
direction of $P_h$ in the photon-proton cm frame (see Fig.~\ref{f:kin}). We have
\begin{align}
\vert \bm{R} \vert &= \frac{1}{2}\, \sqrt{M_h^2 - 2(M_1^2+M_2^2) + (M_1^2-M_2^2)^2/M_h^2} 
\; ,  \nn \\
\bm{R}_T &= \bm{R} \sin \theta  \; . 
\label{e:Rvect}
\end{align}

As usual in SIDIS, we define also the following kinematic invariants
\begin{equation}
\centering
x = \frac{Q^2}{2\,P\cdot q} \; , 
\quad 
y = \frac{P \cdot q}{P \cdot k} \; ,
\quad
\g = \frac{2 M x}{Q} \; , 
\label{e:invariants}
\end{equation}
\begin{equation}
\centering
z = \frac{P \cdot P_h}{P\cdot q}\equiv z_1 + z_2 \; ,
\quad
\zeta = \frac{z_1 - z_2}{z}  \; ,
 \label{e:zetaz}
\end{equation}
where $z_1, \, z_2,$ are the fractional energies carried by the two final 
hadrons. The invariant $\zeta$ can be shown to be a linear polynomial in 
$\cos \theta$~\cite{Bacchetta:2002ux}.

To leading order in the couplings and leading twist, the differential cross section for the two-hadron SIDIS of 
an unpolarized lepton off a transversely polarized nucleon target contains only two nonvanishing structure functions:
\begin{align}
\lefteqn{\frac{d\sig}{dx \, dy\, d\psi \,dz\, d\phi_R\, d M_{h}^2\,d \cos{\theta}} =}  \nn \\ 
& \quad 
\frac{\alpha^2}{x y\, Q^2}\, \frac{y^2}{2\,(1-\varepsilon)}\,  
\biggl( 1+\frac{\g^2}{2x} \biggr)\,
\Biggl\{ F_{UU}  + |\bm{S}_T|\, \varepsilon\, \sin(\phi_R+\phi_S)\,  F_{UT}^{\sin (\phi_R +\phi_S)}
\Biggr\} \; ,
\label{e:crossmaster}
\end{align}
where $\alpha$ is the fine structure constant and the structure functions $F$ depend 
on $x$, $Q^2$, $z$, $\cos\theta$, and $M_h$. The first and second subscripts of $F$ 
indicate the polarization of beam and target, respectively. Here, the target polarization 
refers to the virtual-photon direction; the conversion to the experimental polarization 
with respect to  the lepton beam is straightforward and given in Ref.~\cite{Diehl:2005pc}. 
The angle $\psi$ is the azimuthal angle of $\ell'$ around the lepton beam axis with respect to an 
arbitrary fixed direction, which in case of a transversely polarized target we choose to be the 
direction of $S$. The corresponding relation between $\psi$ and $\phi_S$ is given in 
Ref.~\cite{Diehl:2005pc}; in DIS kinematics it turns out $d\psi \approx d\phi_S$.  

The ratio $\varepsilon$ of longitudinal and transverse photon flux in Eq.~\eqref{e:crossmaster} is 
given by~\cite{Bacchetta:2006tn}
\begin{align}
\varepsilon &= \frac{1-y - \textstyle{\frac{1}{4}} \g^2 y^2}
                                   {1-y+\textstyle{\frac{1}{2}} y^2 + \textstyle{\frac{1}{4}} \g^2 y^2}  \; ,  
\label{e:epsilon}
\end{align}  
so that the depolarization factors can be written as
\begin{alignat}{3}
\frac{y^2}{2\,(1-\varepsilon)} &= 
\frac{1}{1+\g^2} \left(1-y+\textstyle{\frac{1}{2}} y^2 + \textstyle{\frac{1}{4}} \g^2 y^2 \right)
&&\approx \left(1-y +\textstyle{\frac{1}{2}} y^2\right) &&\equiv A(y) \; , \nn 
\\
\frac{y^2}{2\,(1-\varepsilon)}\,\varepsilon &= 
\frac{1}{1+\g^2} \left(1-y- \textstyle{\frac{1}{4}} \g^2 y^2 \right) 
&&\approx (1-y) && \equiv B(y) \; .
\label{e:depol}
\end{alignat} 
The $\varepsilon$ turns out to be also the ratio between the two depolarization factors: 
$\varepsilon = B(y)/A(y)$. Neglecting target-mass corrections, we will assume that in each 
experimental bin 
\begin{equation}
\centering
\langle A(y) \rangle \approx A(\langle y \rangle)\; , 
\quad
\langle B(y) \rangle \approx B(\langle y \rangle) \; ,
\nn 
\end{equation}
\begin{equation}
\centering
\varepsilon \approx \frac{B(\langle y \rangle)}{A(\langle y \rangle)} \equiv C_y  \; . 
\label{e:cy}
\end{equation}

In the limit $M_h^2 \ll Q^2$, the structure functions can be written as products of PDFs and 
DiFFs~\cite{Bianconi:1999cd,Radici:2001na,Bacchetta:2002ux}\footnote{For some discussion 
of the case $M_h^2 \approx Q^2$, see Ref.~\cite{Zhou:2011ba}}
\begin{align} 
F_{UU} & = x \sum_q e_q^2\, f_1^q(x; Q^2)\, D_1^q\bigl(z,\cos \theta, M_h; Q^2\bigr) \; , 
\label{e:FUU} \\
F_{UT}^{\sin (\phi_R +\phi_S)} &=  \frac{|\bm R| \sin \theta}{M_h}\, x\, 
\sum_q e_q^2\,  h_1^q(x; Q^2)\,H_1^{\open\, q}\bigl(z,\cos \theta, M_h; Q^2\bigr) \; , 
\label{e:FUT}
\end{align}
where $e_q$ is the fractional charge of a parton with flavor $q$. The $D_1^q$ is the 
unpolarized DiFF describing the hadronization of a parton with flavor $q$ into an unpolarized 
hadron pair plus anything else, averaging over the parton polarization. The $H_1^{\open\, q}$ 
is a chiral-odd DiFF describing the correlation between the transverse polarization of the 
fragmenting parton with flavor $q$ and the azimuthal orientation of the plane containing the 
momenta of the detected hadron pair. 

For $M_h\ll Q$, the hadron pair can be assumed to be produced mainly in relative $s$ or $p$ waves, 
suggesting that the DiFFs can be conveniently expanded in partial waves. In the two-hadron 
cm frame, the relevant changes in the kinematics are summarized in Eq.~\eqref{e:Rvect}. From the 
simple relation between $\zeta$ and $\cos \theta$, DiFFs can be expanded in Legendre polynomials 
in $\cos \theta$ as~\cite{Bacchetta:2002ux}
\begin{align} 
D_1^{} &\rightarrow D_{1,ss+pp}^{} + D_{1,sp}^{}\cos\theta + 
D_{1,pp}^{} \frac{1}{4}\,(3\cos^2\theta -1) \; , \nn \\
H_1^{\open} &\rightarrow  H_{1,sp}^{\open} + H_{1,pp}^{\open} \cos\theta \; . 
\label{e:partial}
\end{align}
All the $\cos \theta -$dependent terms disappear after integrating upon $d \cos \theta$; they 
still vanish even if the $\theta$ dependence of the acceptance is not complete but symmetric 
about $\theta = \pi / 2$. Of the remaining terms, the subscript $ss+pp$ refers to the unpolarized 
pair being created in a relative $\Delta L=0$ state, while $sp$ indicates the interference in 
$|\Delta L| = 1$. For simplicity, we will use the notation $D_{1,ss+pp}^{} \equiv D_1$ since no 
ambiguity arises in the following. 

From the cross section of Eq.~\eqref{e:crossmaster}, and by inserting the structure functions of 
Eqs.~\eqref{e:FUU}, \eqref{e:FUT} with the approximation of Eq.~\eqref{e:partial}, we obtain 
the following single-spin asymmetry (SSA)~\cite{Radici:2001na,Bacchetta:2002ux,Bacchetta:2006un}
\begin{align}
\lefteqn{A_{UT}^{\sin (\phi_R^{}+\phi_S^{})\,\sin\theta} (x, z, M_h; Q) =} \nn 
\\ 
&\quad =  \frac{1}{|\bm{S}_T|} \, 
\frac{\frac{8}{\pi} \int d\phi_R^{}\, d\cos\theta\, \sin (\phi_R^{}+\phi_S^{}) \, (d\sig^\uparrow - 
           d\sig^\downarrow)}
         {\int d\phi_R^{}\, d\cos\theta\, (d\sig^\uparrow + d\sig^\downarrow)}  = 
\frac{\frac{4}{\pi}\,\varepsilon  \int d\cos\theta\, F_{UT}^{\sin (\phi_R^{}+\phi_S^{})} }
        {\int d\cos\theta\, F_{UU}^{} } \nn 
\\
&\quad = - C_y \,\frac{|\bm{R} |}{M_h} \, 
\frac{ \sum_q\, e_q^2\, h_1^q(x; Q^2)\, H_{1, sp}^{\open\, q}(z, M_h; Q^2)    } 
        { \sum_q\, e_q^2\, f_1^q(x; Q^2)\, D_{1}^q (z, M_h; Q^2) } \nn \\
&\quad \equiv A_{\mathrm{SIDIS}} \;  .
\label{e:ssa}
\end{align} 

We are interested in the specific case of semi-inclusive production of $\pi^+ \pi^-$ pairs. Then, 
isospin symmetry and charge conjugation suggest~\cite{Bacchetta:2006un,Bacchetta:2011ip}
\begin{gather}
D_{1}^{u} = D_{1}^{\bar{u}} \; , \quad D_{1}^{d} =  D_{1}^{\bar{d}}\; ,  \quad 
D_1^{s} = D_1^{\bar{s}} \; , \label{eq:symmD1} \\
H_{1}^{\open\, u} = - H_{1}^{\open\, d} = - H_{1}^{\open\, \bar{u}} = 
H_{1}^{\open\, \bar{d}} \; , \quad 
H_{1}^{\open\, s} = - H_{1}^{\open\, \bar{s}} = 0 \; .  \label{eq:symmH1}
\end{gather}

For a proton target, the SSA~\eqref{e:ssa} simplifies to~\cite{Bacchetta:2011ip}
\begin{align} 
\lefteqn{A_{\mathrm{SIDIS}}^p (x, z, M_h; Q^2) =} \nn \\
&\quad  - C_y \, \frac{|\bm{R}|}{M_h} H_{1, sp}^{\open\, u}(z, M_h; Q^2) \, 
                            \left[ h_1^{u_v}(x; Q^2) - \textstyle{\frac{1}{4}}  h_1^{d_v}(x; Q^2) \right]  \nn \\
&\quad \times \bigg\{ f_1^{u+\bar{u}}(x; Q^2) \, D_1^u (z, M_h; Q^2) + 
                                 \textstyle{\frac{1}{4}} f_1^{d+\bar{d}}(x; Q^2) \, D_1^d (z, M_h; Q^2)  \nn \\
&\qquad                 + \textstyle{\frac{1}{4}} \, f_1^{s+\bar{s}}(x) \, D_1^{s+\bar{s}} (z, M_h; Q^2) 
                           \bigg\}^{-1}  \; ,  \label{e:ssa-p}
\end{align} 
and for a deuteron target to 
\begin{align} 
\lefteqn{A_{\mathrm{SIDIS}}^D (x, y, z, M_h; Q^2) =} \nn \\
&\quad - C_y \, \frac{3}{4}\, \frac{|\bm{R}|}{M_h} H_{1, sp}^{\open\, u}(z, M_h; Q^2) \, 
                                                \left[ h_1^{u_v}(x; Q^2) + h_1^{d_v}(x; Q^2) \right]   \nn \\
&\quad \times \bigg\{ \left[ f_1^{u+\bar{u}}(x; Q^2) + f_1^{d+\bar{d}}(x; Q^2) \right] \,
                                      \left[ D_1^u (z, M_h; Q^2) + \textstyle{ \frac{1}{4}} \, D_1^d (z, M_h; Q^2) \right]  \nn \\
&\qquad                   + \textstyle{\frac{1}{2}} \, f_1^{s+\bar{s}}(x; Q^2) \, D_1^{s+\bar{s}} (z, M_h; Q^2) 
                          \bigg\}^{-1}     \; ,   \label{e:ssa-D}
\end{align} 
where $h_1^{q_v} \equiv h_1^q - h_1^{\bar{q}}$ and $f_1^{q+\bar{q}} \equiv f_1^q + f_1^{\bar{q}}$. 

Eqs.~\eqref{e:ssa-p} and \eqref{e:ssa-D} contains two sets of unknowns: the transversity $h_1$ 
(in various flavor combinations) and the DiFFs. Before the measurement by the Belle 
collaboration of the angular distribution of two pion pairs produced in $e^+ e^-$ 
annihilations~\cite{Vossen:2011fk}, the only information available on DiFFs were coming 
from model calculations in the context of the spectator 
approximation~\cite{Bianconi:1999uc,Radici:2001na,Bacchetta:2006un} 
(and, recently, also using the NJL-jet model~\cite{Casey:2012ux}). The unpolarized 
$D_1$ was tuned to the Monte Carlo event generator~\cite{Bacchetta:2006un} and the polarized 
$H_{1\, sp}^\open$ compared to the asymmetry measured by the HERMES collaboration 
in SIDIS on proton targets~\cite{Bacchetta:2008wb}, the only available set of experimental 
data at that time~\cite{Airapetian:2008sk}. 

The first analysis of the so-called Artru--Collins asymmetry~\cite{Boer:2003ya} in 
$e^+ e^-$ annihilations by the Belle collaboration made possible a direct extraction of 
$H_{1\, sp}^\open$ for the production of $\pi^+ \pi^-$ pairs. In the 
absence of a measurement of the unpolarized $e^+ e^-$ cross section (planned at Belle 
in the near future), $D_1$ was parametrized to reproduce the two-hadron yield of the 
PYTHIA event generator, which is known to give a good description of data. Combining 
such a parametrization with the fit of the azimuthal asymmetry presented in 
Ref.~\cite{Vossen:2011fk}, it was possible to extract for the first time the 
$H_{1\, sp}^\open$~\cite{Courtoy:2012ry}. 

The knowledge of DiFFs in Eq.~\eqref{e:ssa-p} allowed us to get a glimpse of the combination 
$h_1^{u_v} - h_1^{d_v}/4$ directly from the HERMES data for 
$A_{\mathrm{SIDIS}}^p$~\cite{Bacchetta:2011ip}. The effects produced by evolving DiFFs 
between the HERMES and Belle very different scales were properly included by using 
standard evolution equations in a collinear framework~\cite{Ceccopieri:2007ip} and by 
implementing leading-order (LO) chiral-odd splitting functions in the HOPPET code~\cite{Salam:2008qg}. 
Recently, the COMPASS collaboration has released new data for $A_{\mathrm{SIDIS}}^p$ 
on a proton target and for $A_{\mathrm{SIDIS}}^D$ on a deuteron target, with higher 
statistics and wider kinematic coverage~\cite{Adolph:2012nw}. Thus, the combination of 
SSA of Eqs.~\eqref{e:ssa-p} and \eqref{e:ssa-D} makes it possible to separately parametrize 
each valence flavor of the transversity distribution, which we present here
for the first time (see also Ref.~\cite{Elia:2012} for a first attempt to obtain a 
flavor separation point by point). 

In the collinear framework, the dependence of the SSA on the momentum fraction $x$ gets 
factorized from the dependence on $(z, M_h)$, as it is evident in Eqs.~\eqref{e:ssa-p} and 
\eqref{e:ssa-D}. This suggests that the dependence of the SSA on $x$ comes 
only from the involved PDFs. Therefore, it is more convenient to study it by integrating the 
$z$- and $M_h$-dependence of DiFFs. Then, the actual combinations of transversity used 
in the analysis are, for the proton, 
\begin{equation} 
\begin{split}
x\, h_1^{p} &(x; Q^2) \equiv x \, h_1^{u_v}(x; Q^2) - {\textstyle \frac{1}{4}}\, x h_1^{d_v}(x; Q^2) \\
&= -\frac{ A^p_{\text{SIDIS}} (x; Q^2)  }{C_y\, n_u^{\uparrow}(Q^2)}   \\ 
&\quad  \times \biggl[ 
n_u (Q^2)\, x f_1^{u+\bar{u}}(x; Q^2) + \textstyle{\frac{1}{4}}\, n_d (Q^2)\, x f_1^{d+\bar{d}}(x; Q^2)
+ \textstyle{\frac{1}{4}}\, n_s (Q^2)\, x f_1^{s+\bar{s}}(x; Q^2)  \biggr]  \; ,
\label{e:h1p} 
\end{split}   
\end{equation} 
and, for the deuteron, 
\begin{equation} 
\begin{split} 
 x\, h_1^{D} &(x; Q^2) \equiv x \, h_1^{u_v}(x; Q^2)+ x h_1^{d_v}(x; Q^2)   \\
 &=- \frac{A^D_{\text{SIDIS}}(x; Q^2)}{C_y \, n_u^{\uparrow}(Q^2)} \frac{4}{3}  \\
&\quad \times \, \biggl[ \biggl(
x f_1^{u+\bar u}(x; Q^2)+ x f_1^{d+\bar d}(x; Q^2) \biggr)  \biggl( n_u(Q^2)+ \frac{n_d(Q^2)}{4} \biggr)
+ \frac{n_s(Q^2)}{2} \, x f_1^{s+\bar s}(x; Q^2) \biggr]  \; ,
\label{e:h1D}
\end{split} 
\end{equation} 
where
\begin{align} 
n_q(Q^2) &= \int_{z_{\text{min}}}^{z_{\text{max}}} \int_{M_{h\, \text{min}}}^{M_{h\, \text{max}}} 
dz \, dM_h \, D_1^q (z, M_h; Q^2)  \; , \label{e:nq}  \\
n_q^\uparrow (Q^2) &= \int_{z_{\text{min}}}^{z_{\text{max}}} \int_{M_{h\, \text{min}}}^{M_{h\, \text{max}}} 
dz \, dM_h \, \frac{|\bm{R}|}{M_h}\, H_{1\, sp}^{\open\, q}(z,M_h; Q^2) \; .
\label{e:nqperp}
\end{align} 
Using the unpolarized PDFs from the MSTW08LO set~\cite{Martin:2009iq}, early explorations were 
presented in Ref.~\cite{Courtoy:2012in}.


\section{Fitting procedure}
\label{s:h1param}

Here, we describe our fitting procedure to obtain the valence transversity
distribution functions for up and down quarks. We discuss first the choice of the 
functional form. 

The main theoretical constraint we have is Soffer's inequality~\cite{Soffer:1995ww} 
(see also \cite{Goldstein:1995ek}) 
\begin{equation} 
2|h_1^q(x; Q^2)| \leq | f_1^q(x; Q^2) + g_1^q(x; Q^2)| \equiv 2\,\mbox{\small SB}^q(x; Q^2) \; .
\label{e:soffer}
\end{equation} 
We impose this condition by multiplying the functional form by the corresponding Soffer
bound at the starting scale of the parameterization. If the Soffer bound is fulfilled at some 
initial $Q_0^2$, it will hold also at higher $Q^2 \geq Q_0^2$~\cite{Bourrely:1998bx,Vogelsang:1997ak}. 
The implementation of the Soffer bound depends on the choice of the unpolarized and helicity PDFs. 
We use the MSTW08 set~\cite{Martin:2009iq} for the unpolarized PDF, combined to the 
DSSV parameterization~\cite{deFlorian:2009vb} for the helicity distribution, at the 
scale of $Q_0^2=1$ GeV$^2$. Our analysis was carried out at LO in $\alpha_S$. 
To be as consistent as possible, we decided to use the MSTW08LO set 
for $f_1$ and the DSSV set for $g_1$, even if the DSSV fit provides only a NLO
parametrization of $g_1$.
For convenience, in App.~\ref{s:app} we list the explicit form of $\mbox{\small SB}^q$. 
The result for the Soffer bound is affected by an error coming mainly from the 
uncertainty in the knowledge of the helicity PDF $g_1$. We checked that at the explored 
hard scales this error is much smaller than the experimental errors on $A_{\mathrm{SIDIS}}$ 
data and the statistical error on the DiFF parametrization; hence, we will neglect it. 

The Soffer bound is valid for each quark and antiquark. Since we need to parametrize 
the transversity valence combinations for up and down quarks, we have necessarily to 
constrain it by taking the sum of Soffer bounds for both quarks and antiquarks. This likely 
leads to a loose bound, especially at low $x$. In particular, due to the divergent behavior 
of PDFs the ``valence'' Soffer bound is not even integrable in the range $x \in [0,1]$,
which would result in a divergent tensor charge. Given the chosen analytical form of
the PDFs at low $x$, the $\mbox{\small SB}^q$ has to be multiplied by at least 
$x^{0.16276}$.

Based on the above considerations, we adopted the following functional form
for the valence transversity distributions at $Q_0^2 = 1$ GeV$^2$:
\begin{equation} 
x\, h_1^{q_V}(x; Q_0^2)=
\tanh \Bigl[ x^{1/2} \, \bigl( A_q+B_q\, x+ C_q\, x^2+D_q\, x^3\bigr)\Bigr]\, 
\Bigl[ x\, \mbox{\small SB}^q(x; Q_0^2)+x\, \mbox{\small SB}^{\bar q}(x; Q_0^2)\Bigr] \; .
\label{e:funct_form}
\end{equation} 
The hyperbolic tangent is such that the Soffer bound is always fulfilled. 
The functional form is very flexible and can contain up to three nodes. 
The low-$x$ behavior is however determined by the $x^{1/2}$ term, which is imposed 
by hand. Present fixed-target data do not allow to constrain it. 

The fit, and in particular the error analysis, 
was carried out in two different ways: using the
standard Hessian method and using a Monte Carlo approach. As usual, we remind
the reader that both methods are suitable to estimate the errors of
statistical nature only, assuming a specific choice of the theoretical
function. Care must be taken especially when using error bands outside the
region where data exist. 

The standard fitting procedure consists in minimizing the usual $\chi^2$ function, defined as
\begin{equation}
\chi^2(\{p\}) = \sum_{i} 
\frac{\Bigl( x_i\, h_{1, \mbox{\tiny data}}^{p/D}(x_i; Q^2_i) - x_i\, h_{1, \mbox{\tiny theo}}^{p/D}(x_i, Q^2_i;\{p\}) \Bigr)^2}
        {\Bigl( \Delta h_{1, \mbox{\tiny data}}^{p/D}(x_i; Q^2_i)\Bigr)^2}  \; , 
\label{e:stand_chi2}
\end{equation}
where the sum runs over the experimental points, the expressions for $x\, h_1^{p/D}$ 
are listed in Eqs.~\eqref{e:h1p} and \eqref{e:h1D}, and $\{p\}$ denotes the vector of parameters.
The evolution of the functional form~\eqref{e:funct_form} to the values $Q^2_i$ for each data bin 
has been implemented using the HOPPET code~\cite{Salam:2008qg}, set to the $\overline{\mbox{MS}}$ renormalization scheme and modified to include also the LO chiral-odd splitting functions needed for 
transversity evolution. The input value for the running coupling constant at $Q_0^2=1$ GeV$^2$ 
is chosen to be the best-fit value of the MSTW08LO set, i.e. 
$\alpha_S^{\mbox{\tiny LO}}(Q_0^2) = 0.13939$. It is true for all the evolved quantities of our 
analysis, including the Soffer bound. The minimization has been carried out using the MINUIT code 
and led to a vector of best-fit parameters, $\{ p_0\}$ (and a covariance matrix). 

The standard method allows to compute the errors on any theoretical quantity
under the assumption that the parameter dependence of $\chi^2$ 
can be approximated by a
quadratic expansion around the minimum, and the parameter dependence 
of the theoretical quantity can be approximated by a
linear expansion around the minimum.

For the standard method, the error on the extracted transversity was estimated
using the formula
\begin{equation} 
\Bigl( \Delta h_1 (x, Q^2; \{ p\})\Bigr)^2 =
\sum_{i,j}^{N_p}\, \frac{\partial h_{1,\mbox{\tiny theo}} (x, Q^2;\{ p\})}{\partial p_i}\Bigr\vert_{\{ p_0\}} 
\, \text{Cov}_{ij}\,
\frac{\partial h_{1,\mbox{\tiny theo}} (x,Q^2 ; \{p\})}{\partial p_j}\Bigr\vert_{\{ p_0\}} \; ,
\label{e:Hesserr}
\end{equation} 
where $N_P$ is the number of parameters. The covariance matrix $\text{Cov}_{ij}$ 
has been obtained using the condition $\Delta \chi^2 =1$. Therefore, 
within the limits of applicability of the standard approach, the obtained error band
corresponds to the $1 \sigma$ or 68\% confidence level. In typical PDF global fits, 
often the value of $\Delta \chi^2$ is increased of one or even two orders of
magnitude, with a corresponding increase in the error estimate (see, e.g.,
\cite{deFlorian:2009vb}). In the present analysis, we find no need of such an increase, 
as demonstrated by the agreement with the Monte Carlo approach.

The Monte Carlo approach does not rely on the assumptions of a quadratic
dependence of $\chi^2$ and a linear expansion of the theoretical quantity 
around $\{ p_0\}$, respectively. In our case, the need of such an approach is essential 
whenever the minimization pushes the theoretical functions towards their upper or lower
bounds, where it is not possible to assume a simple linear expansion in the
parameters. 

For the implementation of this approach, we took inspiration from the work of
the NNPDF collaboration (see, e.g., \cite{Forte:2002fg,Ball:2008by,Ball:2010de}), 
although our results are not based on a neural-network fit. The approach consists in 
creating $N$ replicas of the data points. In each replica (denoted by the index $r$), 
each data point $i$ is shifted by a Gaussian noise with the same variance as the 
measurement. Each replica, therefore, represents a possible outcome of an 
independent experimental measurement, which we denote by $h_{1, r}^{p/D}(x_i; Q^2_i)$. 
The number of replicas is chosen so that the mean and standard deviation of the 
set of replicas accurately reproduces the original data points. In our case, we have 
found that 100 replicas are sufficient.

The standard minimization procedure is applied to each replica separately, by
minimizing the following error function\footnote{Note that the error for each replica is taken to be equal
  to the error on the original data points. This is consistent with the fact
  that the variance of the $N$ replicas should reproduce the variance of the
  original data points.} 
\begin{equation}
E_r^2(\{p\})=\sum_{i} 
\frac{\Bigl(x_i\,  h_{1, r}^{p/D}(x_i; Q^2_i) - x_i\, h_{1, \mbox{\tiny theo}}^{p/D}(x_i, Q^2_i;\{p\}) \Bigr)^2}
        {\Bigl(\Delta h_{1, \mbox{\tiny data}}^{p/D}(x_i; Q^2_i)\Bigr)^2}  \; , 
\label{e:MC_chi2}
\end{equation}
resulting in $N$ different vectors of best-fit parameter values, $\{ p_{0r}\},\; r=1,\ldots N$. 
These parameter vectors can be used to produce $N$ values for any theoretical quantity. 
The $N$ theoretical outcomes can have any distribution, not necessarily Gaussian. 
For non-Gaussian distributions, the $1 \sigma$ confidence interval is in general different 
from the 68\% interval. Both of them can be easily computed from the $N$ theoretical outcomes. 
For instance, for the 68\% interval we simply take for each experimental point $i$ the $N$ values 
and we reject the largest and the lowest 16\% of them.

Although the minimization is performed on the function defined in Eq.~\eqref{e:MC_chi2}, 
the agreement of the $N$ theoretical outcomes with the original data is better expressed in 
terms of the original $\chi^2$ function defined in Eq.~\eqref{e:stand_chi2}, i.e. with respect 
to the original data set without the Gaussian noise. If the model is able to give a good
description of the data, the distribution of the $N$ values of $\chi^2$/d.o.f. should be peaked at 
around one. In real situations, the rigidity of the model shifts the position of the peak to 
higher values of $\chi^2$/d.o.f..

In our case, we determined $N = 100$ best-fit parameter vectors and we used them 
to produce 100 curves for the up and down valence transversity. Each one of the resulting 
curves respects the Soffer bound by construction. The results are discussed in the next section.


\section{Results and discussion}
\label{s:out}

In the following, we discuss the results obtained by fitting the expressions of 
Eqs.~\eqref{e:h1p} and \eqref{e:h1D} when inserting the HERMES and COMPASS 
measurements for the single-spin asymmetries $A_{\text{SIDIS}}^p$ and 
$A_{\text{SIDIS}}^D$ on $\pi^+ \pi^-$ SIDIS production off transversely polarized 
proton and deuteron targets, respectively. By combining the two fits, we can 
determine for each valence flavor $u_v$ and $d_v$ the vector of fitting parameters 
that gives the corresponding transversity distribution, according to Eq.~\eqref{e:funct_form}.

\begin{table}
  \centering
  \small
  \begin{tabular}{|ccccc|}
  \hline
    &HERMES &data & & \\
   \hline
    $x$       &	$y$	& $Q^2 {[\mbox{GeV}^2]}$ & $A_{\text{SIDIS}}$  &    $h_1^{u_v}- h_1^{d_v}/4$ \\
    \hline
    $0.033$  &	$0.734$   & $1.232$     & \,$0.015\pm 0.010$     & $ 0.086 \pm 0.061$\\
    $0.047$  &	$0.659$	 &  $1.604$ 	& \,$0.002\pm 0.011$	 & $0.010 \pm 0.054$\\
    $0.068$  &	$0.630$	 &  $2.214$ 	& \,$0.035\pm 0.011$	 & $0.167 \pm 0.069$ \\
    $0.133$  &	$0.592$	 &  $4.031$ 	& \,$0.020\pm 0.010$	 & $0.092 \pm 0.054$ \\
    \hline
    \hline
     &  COMPASS & proton & data &\\
     \hline
       $x$       &	& $Q^2{[\mbox{GeV}^2]}$ & $A_{\text{SIDIS}}$  & $h_1^{u_v}- h_1^{d_v}/4$ 	\\
    \hline
    $0.0065$ 	& 	&	$1.232$		& $0.026\pm 0.030$  & $0.10\pm 0.12$    \\	
    $0.0105$ 	&	&  	$1.476$ 	& $0.010\pm 0.016$  & $0.038\pm 0.059$  \\	
    $0.0164$ 	&	&  	$1.744$ 	& $0.015\pm 0.013$  & $0.057 \pm 0.049$ \\	
    $0.1330$ 	&	&  	$2.094$ 	& $0.008\pm 0.010$  & $0.031 \pm 0.039$ \\	
    $0.0398$	&	&  	$2.802$		& $0.027\pm 0.011$  & $0.107 \pm 0.049$	\\	
    $0.0626$	&	&	$4.342$		& $0.029\pm 0.014$  & $0.118 \pm 0.060$	\\	
    $0.1006$	&	&	$6.854$		& $0.051\pm 0.016$  & $0.208 \pm 0.079$ \\	
    $0.1613$	&	&	$10.72$		& $0.108\pm 0.023$  & $0.42 \pm 0.12$ \\	
    $0.2801$	&	&	$21.98$		& $0.080\pm 0.033$  & $0.24 \pm 0.11$ \\	
 \hline
 \hline       
       & COMPASS &deuteron &data &\\
   \hline
	  $x$       &	& $Q^2{[\mbox{GeV}^2]}$ & $A_{\text{SIDIS}}$ & $h_1^{u_v}+h_1^{d_v}$ 	\\
    \hline
         $0.0064$	&	&       $1.253$	&	$0.005\pm 0.024$ & $0.05 \pm 0.24$	\\	
	 $0.0105$	&	&	$1.508$	&	$-0.004\pm 0.012$ & $-0.04 \pm 0.12$	\\
	 $0.0163$	&	&	$1.792$	&	$ 0.028\pm 0.010$ & $0.28 \pm 0.11$	\\
	 $0.0253$	&	&	$2.266$	&	$-0.005\pm 0.009$ & $-0.051 \pm 0.094$	\\
	 $0.0396$	&	&	$3.350$	&	$ 0.006\pm 0.011$ & $0.06 \pm 0.12$	\\
	 $0.0623$	&	&	$5.406$	&	$-0.006\pm 0.014$ & $-0.06 \pm 0.14$	\\
	 $0.0996$	&	&	$8.890$	&	$-0.029\pm 0.019$ & $-0.30 \pm 0.20$	\\
	 $0.1597$	&	&	$15.65$	&	$-0.017\pm 0.030$ & $-0.16 \pm 0.28$	\\
	  $0.2801$	&	&	$33.22$	&	$ 0.078\pm 0.054$ & $0.50 \pm 0.36$	\\
     \hline
  \end{tabular}
  \caption{
HERMES data for $\pi^+ \pi^-$ production in SIDIS off a transversly polarized 
proton~\cite{Airapetian:2008sk} and COMPASS data for the same process off a transversly 
polarized proton and deuteron~\cite{Adolph:2012nw}. The last column shows the combinations 
of valence transversities obtained using Eqs.~\eqref{e:h1p} and \eqref{e:h1D}.}
\label{t:data}
\end{table}

As discussed in the previous section, the error analysis has been performed in two ways: 
using the standard Hessian method summarized in Eqs.~\eqref{e:stand_chi2} and 
\eqref{e:Hesserr}, or the Monte Carlo approach by fitting $N = 100$ replicas of the 
experimental points according to Eq.~\eqref{e:MC_chi2}. For each strategy, we 
explored three different scenarios in the parametrization~\eqref{e:funct_form} of the 
transversity distribution:
\begin{itemize}
\item the {\it rigid} scenario, described by the choice  $C_u=C_d=D_u=D_d=0$, i.e. with only 
4 free parameters;
\item the {\it flexible} scenario, with $D_u=D_d=0$ (6 free parameters); 
\item the {\it extra-flexible} scenario, with all 8 free parameters.
\end{itemize}


\subsection{Experimental data}
\label{s:datain}

In Tab.~\ref{t:data}, we list the data for the asymmetries $A_{\text{SIDIS}}^p$ and 
$A_{\text{SIDIS}}^D$ of Eqs.~\eqref{e:ssa-p} and \eqref{e:ssa-D}, as they were  measured 
by the HERMES~\cite{Airapetian:2008sk} and COMPASS~\cite{Adolph:2012nw} collaborations 
in the $\pi^+ \pi^-$ SIDIS production off transversely polarized proton and deuteron targets, 
respectively. The first three columns indicate the average values of the corresponding 
kinematic variables in each experimental bin. The indicated errors include statistical and 
systematic contributions added in quadrature. The depolarization factor $C_y$ in the expression of the 
asymmetries depends on the average $y$ according to Eq.~\eqref{e:cy}. In its analysis, the 
COMPASS collaboration already divided the $C_y$ factor out of the measured cross section; 
as such, the SSA does no longer depend on $y$ and, correspondingly, there are no $y$ 
values in the second column of Tab.~\ref{t:data} for COMPASS. Consistently, we have used 
$C_y = 1$ when fitting the COMPASS data.

\begin{table}
  \centering
  \begin{tabular}{|r r r r r |}
 \hline
  &  HERMES & range & for proton &\\
 \hline
    $Q^2$ 	[GeV$^2$]	& $n_u$		&   $ n_d$		& 	  $ n_s$     &  $n_u^{\uparrow}$ \\
 \hline
     $1.232$    		& $0.607$		&	$0.614$		& 	$0.393$	 &	$-0.157\pm  0.037$	\\
     $1.604$		& $0.589$		&	$0.595$		&	$0.380$	 &	$-0.152\pm  0.037$		\\
     $2.214$		& $0.569$		&     $0.575$ 		&	$0.365$	 & 	$-0.146\pm  0.037$	\\
     $4.031$		& $0.536$		&     $0.542$		&  	$0.341$	 &    $-0.137\pm  0.037$ 	\\
 \hline
 \hline
      &  COMPASS & range & for proton &\\
  \hline
    $Q^2$ 	[GeV$^2$]	& $n_u$		&   $ n_d$		& 	  $ n_s$     &  $n_u^{\uparrow}$ \\
  \hline
     $1.232$	&  	$0.897$	&  	$ 0.906$	&  	$  0.580$ 	&  	$ -0.183\pm  0.031$\\
     $1.476$	&      $0.876$	&  	$ 0.885$	&  	$ 0.565$ 	&  	$-0.178 \pm  0.031$\\   
     $1.744$	&      $0.858$	&  	$  0.867$	&  	$  0.552$	&  	$-0.175  \pm  0.031$\\
     $2.094$	&      $0.840$	&  	$  0.849$ 	&  	$ 0.539$ 	&  	$-0.171   \pm  0.031$\\
     $2.802$	&      $0.813$	&  	$  0.822$	&  	$ 0.520$ 	&  	$-0.165   \pm  0.031$\\  
     $4.342$	&       $0.776$	&  	$  0.785$	&  	$ 0.494$ 	&  	$-0.158   \pm  0.031$\\  
     $6.854$	&       $0.742$	&  	$  0.751$	&  	$ 0.471$	&  	$-0.151   \pm  0.031$\\ 
     $10.720$	&       $0.713$	&  	$  0.721$	&  	$ 0.451$	&  	$ -0.145  \pm  0.031$\\   
     $21.985$ 	&       $0.671$	&  	$   0.679$	&  	$ 0.422$ 	&  	$ -0.136  \pm  0.031$\\
  \hline
  \hline
      &  COMPASS & range & for deuteron &\\
   \hline
    $Q^2$ 	[GeV$^2$]	& $n_u$		&   $ n_d$		& 	  $ n_s$     &  $n_u^{\uparrow}$ \\
   \hline
   $1.253$		&  	$0.895$	&  	$ 0.904$	&  	$0.578$	&  	$ -0.182\pm  0.031$\\     
   $1.508$		&  	$0.874$	&  	$0.883$	&  	$0.563$ 	&  	$-0.178 \pm  0.031$\\   
   $1.792$		&  	$0.855$	&  	$0.865$	&  	$0.550$	&  	$ -0.174\pm  0.031$\\    
   $2.266$		&  	$0.832$ 	&  	$0.841$	&  	$0.534$	&  	$-0.169 \pm  0.031$\\    
   $3.350$		&  	$0.797$	&  	$0.806$	&  	$0.509$	&  	$-0.162  \pm  0.031$\\
   $5.406$		&  	$0.759$	&  	$0.768$	&  	$0.483$ 	&  	$-0.154 \pm  0.031$\\   
   $8.890$		&  	$0.725$	&  	$0.733$	&  	$0.459$	&  	$-0.147 \pm  0.031$\\   
   $15.652$	&  	$0.690$	&  	$0.698$	&  	$0.435$	&  	$ -0.140\pm  0.031$\\    
   $33.219$	&  	$ 0.650$	&  	$0.657$	&  	$0.408$	&  	$ -0.132 \pm  0.031$\\ 
  \hline
 \end{tabular}
 \caption{The integrated DiFFs according to Eqs.~\eqref{e:nq} and \eqref{e:nqperp}. The 
 error has been computed at the average $Q^2$ for each indicated experimental bin.}
  \label{t:nq}
\end{table}

The last column in Tab.~\ref{t:data} contains the values of the combination in 
Eq.~\eqref{e:h1p} for the proton target, and of Eq.~\eqref{e:h1D} for the deuteron target, 
when the corresponding experimental values for the SSA are inserted in 
$A_{\text{SIDIS}}^p$ and $A_{\text{SIDIS}}^D$, respectively. As already anticipated in 
the previous sections, for the unpolarized PDFs we adopted the MSTW08LO set~\cite{Martin:2009iq}. 
The remaining ingredients in Eqs.~\eqref{e:h1p} and \eqref{e:h1D} are the 
$n_q$ and $n_q^\uparrow$ defined in Eqs.~\eqref{e:nq} and \eqref{e:nqperp}, 
where the DiFFs $D_1^q$ and $H_{1\, sp}^{\open\, q}$ are parametrized as 
in Ref.~\cite{Courtoy:2012ry}. The integrals are evaluated according to the appropriate 
experimental cuts: $0.2<z<1$ and $0.5\text{ GeV}<M_h<1$ GeV for HERMES,
$0.2<z<1$ and $0.29\text{ GeV}<M_h<1.29$ GeV for COMPASS. In Tab.~\ref{t:nq}, 
the results are given for the relevant flavors at the average scales $Q^2_i$ for each experimental 
bin $i$. The statistical error is indicated only for $n_u^\uparrow$, since the 
large statistics achievable in the Monte Carlo simulation of the unpolarized 
$e^+ e^- \to (\pi^+ \pi^- ) X$ cross section makes the error of $n_q$ negligible.

\begin{table}
  \centering
  \begin{tabular}{|c||c|c|c|}
  \hline
  \multicolumn{4}{|c|}{{\it Rigid} scenario} \\
  \hline
  \hline 
          &  up   &     down   &  $\chi^2/$d.o.f.   \\
  \hline
  $A$    &    $0.76  \pm 0.35$   &  $2.3  \pm 2.7$   &   $22.2/18 = 1.23$   \\
  $B$	    &    $0.5     \pm 2.0$     &  $-81  \pm 69$ &    \\
  \hline
  \hline
  \multicolumn{4}{|c|}{{\it Flexible} scenario} \\
  \hline
  \hline
          &    up    &	   down &    $\chi^2/$d.o.f.    \\
  \hline
  $A$	   &     $1.41    \pm 0.62$     &     $-0.5     \pm 6.8$    &   $17.9/16 = 1.12$   \\
  $B$   &     $-11      \pm 10$        &     $104	  \pm 413$   &     \\
   $C$  &    $35   \pm 35$	      &   $(-22   \pm 54) \times 10^{2}$ &    \\ 
   \hline
   \hline
  \multicolumn{4}{|c|}{{\it Extra-flexible} scenario} \\
  \hline
  \hline
        &   up	     &     down &    $\chi^2/$d.o.f.     \\
   \hline
  $A$    &  $1.79    \pm 0.53$    &     $2.6  \pm 5.0$  &    $17.6/14 = 1.26$    \\
  $B$    &  $-24.7   \pm 8.7$      &     $-239  \pm 352$ &    \\
  $C$    &  $136	  \pm 53$    &    $(82  \pm 99)  \times 10^{2}$ &    \\ 
  $D$    &  $-183  \pm 101$   &  $(-9.2   \pm 10)   \times 10^{4}$ &   \\ 
   \hline
  \end{tabular}
  \caption{Best-fit parameters and $\chi^2$ values obtained in the standard approach
    for the three scenarios described in the text and based on Eq.~\eqref{e:funct_form}.}
  \label{t:param}
\end{table}

In the standard Hessian method, the best fit parameters and their $1\sigma$
error (corresponding to $\Delta \chi^2 =1$) at the initial scale $Q_0^2=1$ GeV$^2$ 
are given in Tab.~\ref{t:param}. The $\chi^2/$d.o.f. is $1.23$ for
the {\it rigid} scenario, $1.12$ for the {\it flexible} scenario, and $1.26$ for the 
{\it extra-flexible} scenario. For the 100 replicas, the average $\chi^2/$d.o.f. 
are 1.35, 1.56
and  1.86, respectively.

\begin{figure}[tb]
  \centering
  \includegraphics[width=0.47\textwidth]{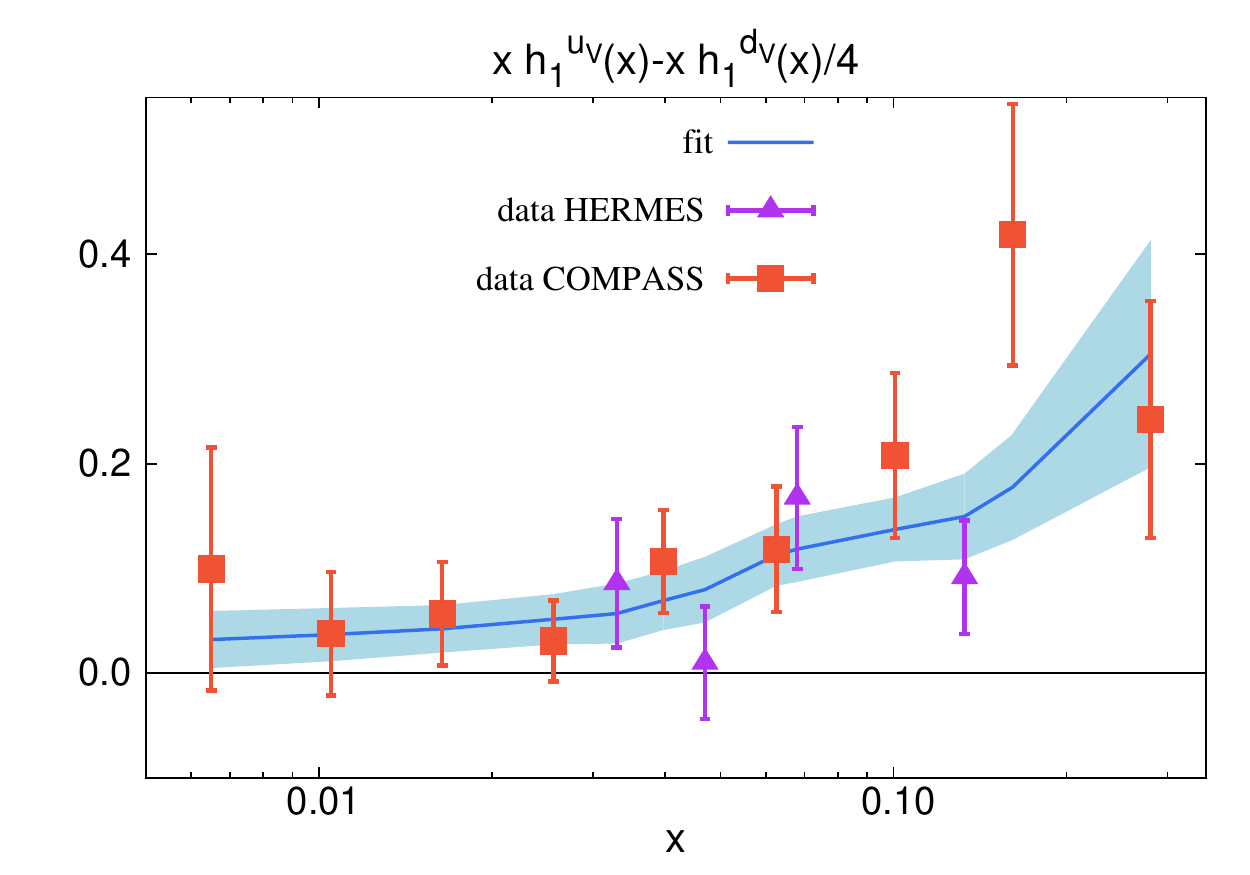}
   \includegraphics[width=0.47\textwidth]{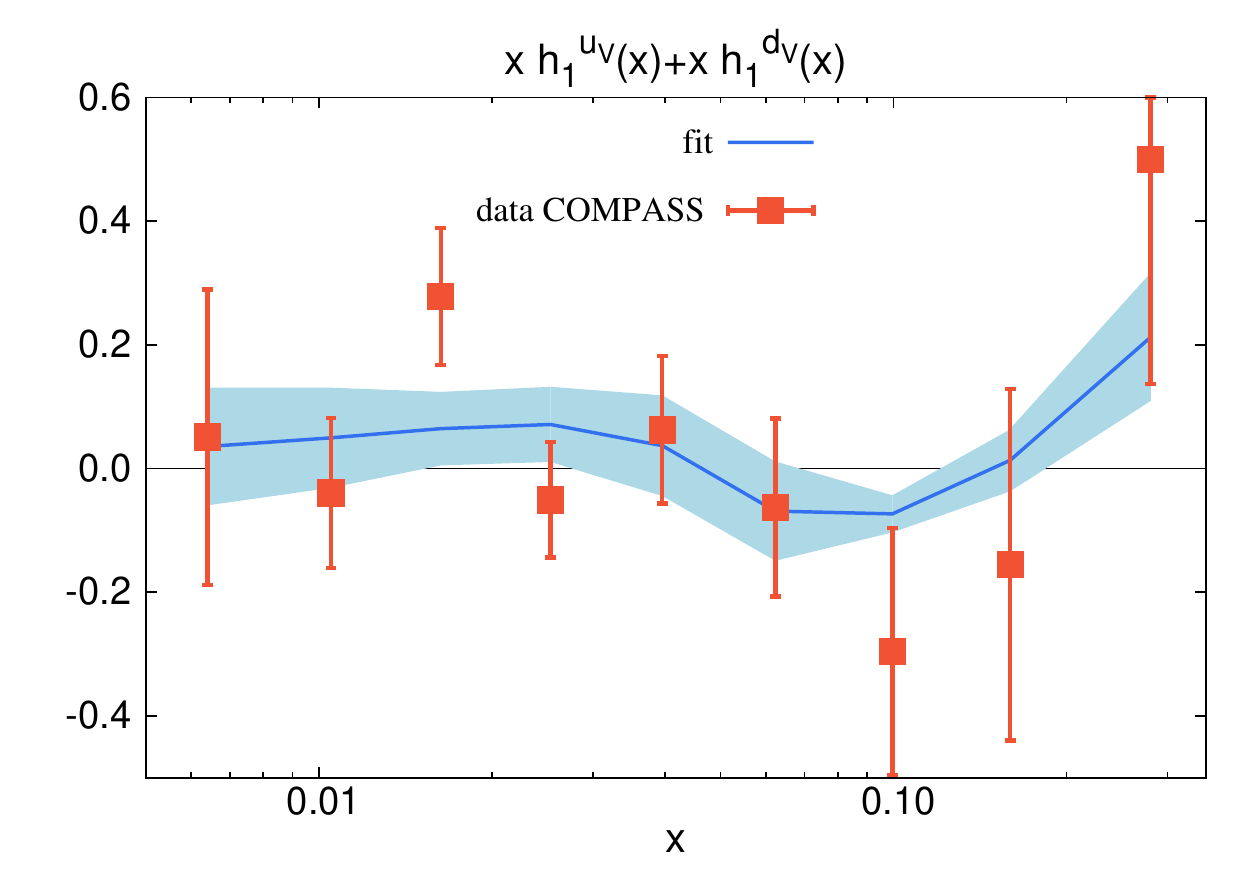}
  \caption{The combinations of Eq.~(\ref{e:h1p}), left panel, and
    Eq.~(\ref{e:h1D}), right panel. The squares and triangles are obtained from 
    the COMPASS and HERMES data, respectively (the values are indicated 
    in the last column of Tab.~\ref{t:data}). The thick solid line indicates the central value 
    of the best-fit result in the standard approach with the {\it flexible} scenario 
    (see text). The error band is the outcome of the merging of all the straight lines 
    connecting the statistical error bars of the fit for each experimental point.} 
  \label{f:xh1stand}
\end{figure}


\subsection{Fitting results}
\label{s:plots}

In Fig.~\ref{f:xh1stand}, the points with error bars represent the transversity 
combinations for proton (left panel) and deuteron target (right panel) 
quoted in the last column of Tab.~\ref{t:data}. The central value of our best-fit
result in the standard approach with the {\it flexible} scenario is given by the 
thick solid line, and it is in good agreement with the data. The error band is 
the outcome of the merging of all the straight lines connecting the statistical 
error bars of the fit for each experimental point. The other scenarios do not show significant 
qualitative differences in the range where data exist.

\begin{figure}[tb]
  \centering
  \includegraphics[width=0.47\textwidth]{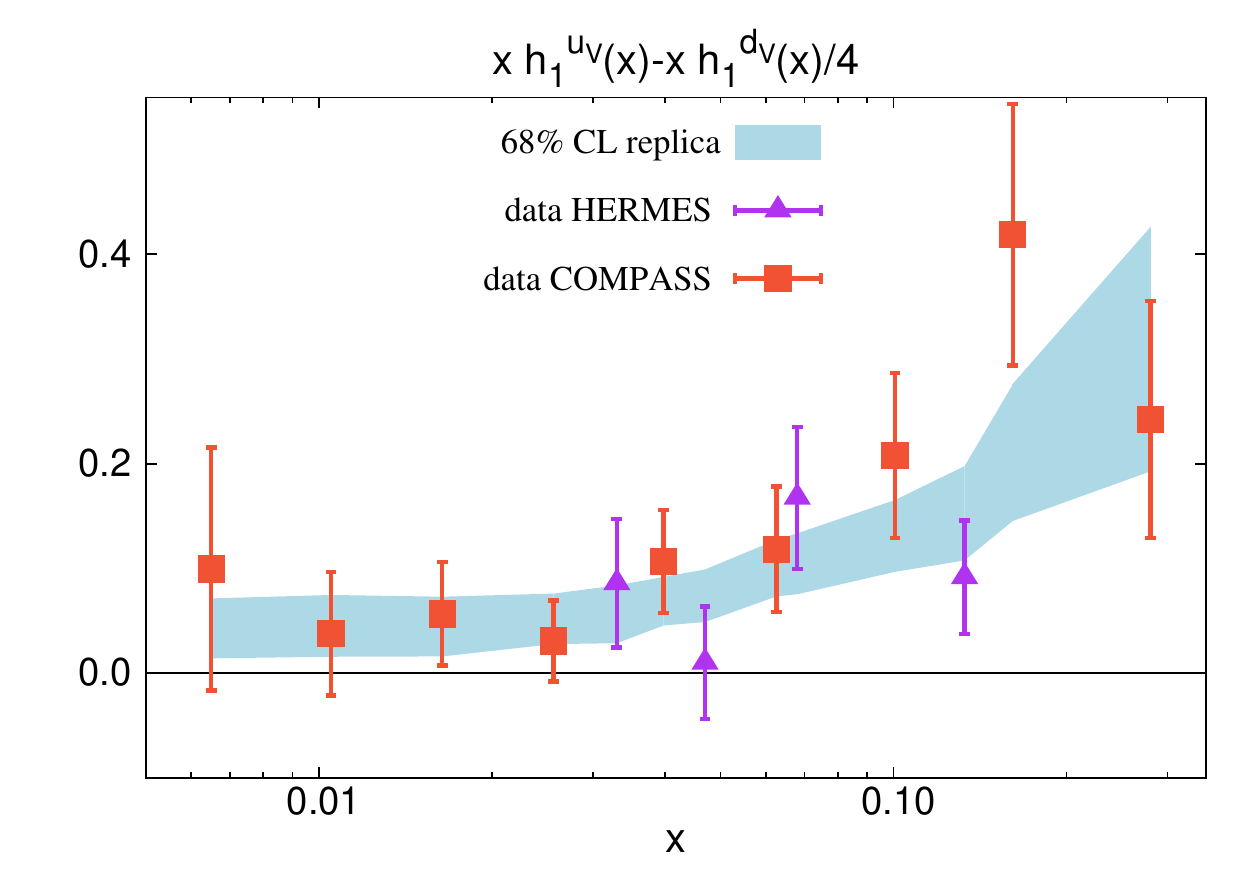}
   \includegraphics[width=0.47\textwidth]{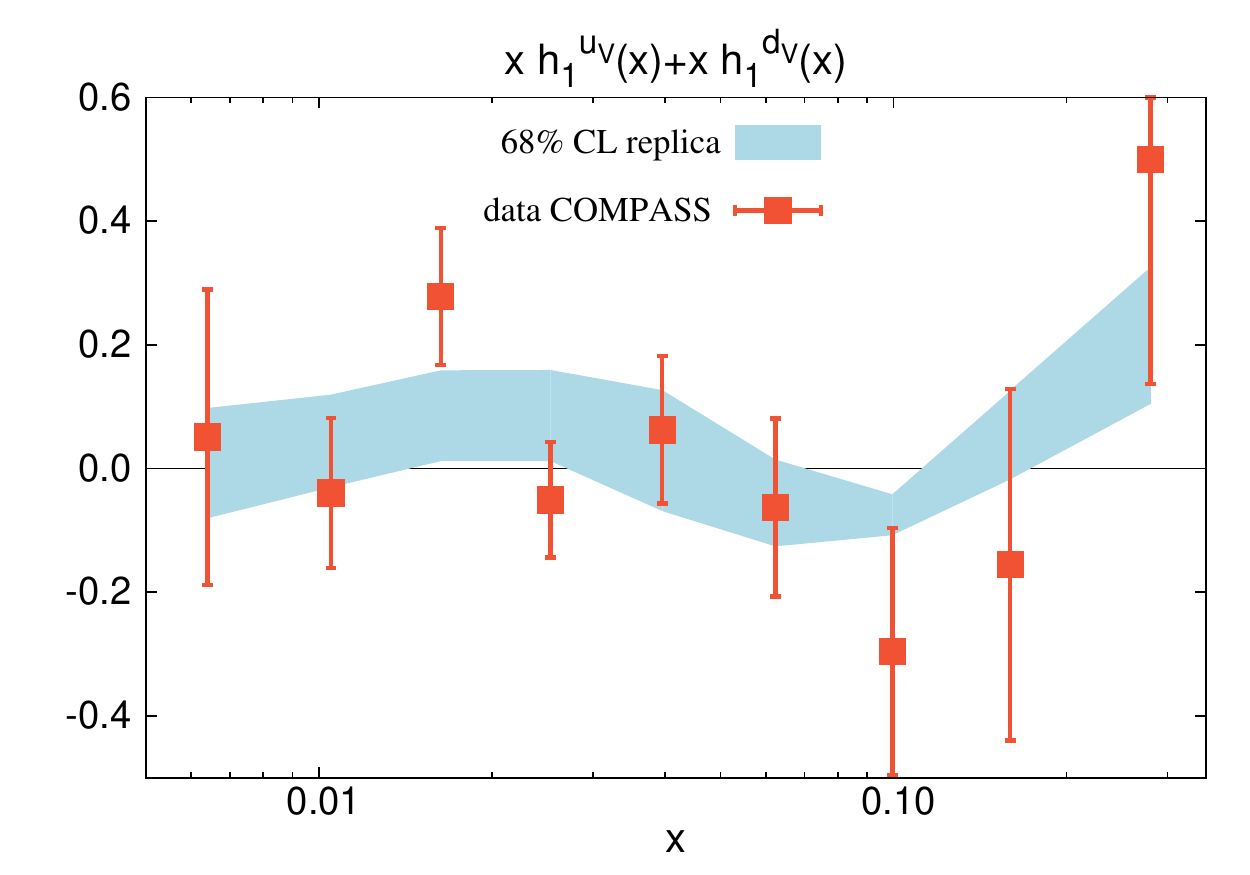}
  \caption{Same observables and data symbols as in the previous figure. The uncertainty 
  band represents in the Monte Carlo approach the selected $68\%$ of all fitting replicas 
  (see text).}
  \label{f:xh1replica}
\end{figure}

In Fig.~\ref{f:xh1replica}, we show the same comparison in the same conditions as 
in the previous figure, but for the Monte Carlo approach. The band now represents 
the result of the $68\%$ of all replicas, obtained by rejecting the largest $16\%$ and 
the lowest $16\%$ of the replicas' values in each $x$ point. We observe no substantial 
difference between the standard and Monte Carlo approaches.

\begin{figure}
  \centering
     \includegraphics[width=0.49\textwidth]{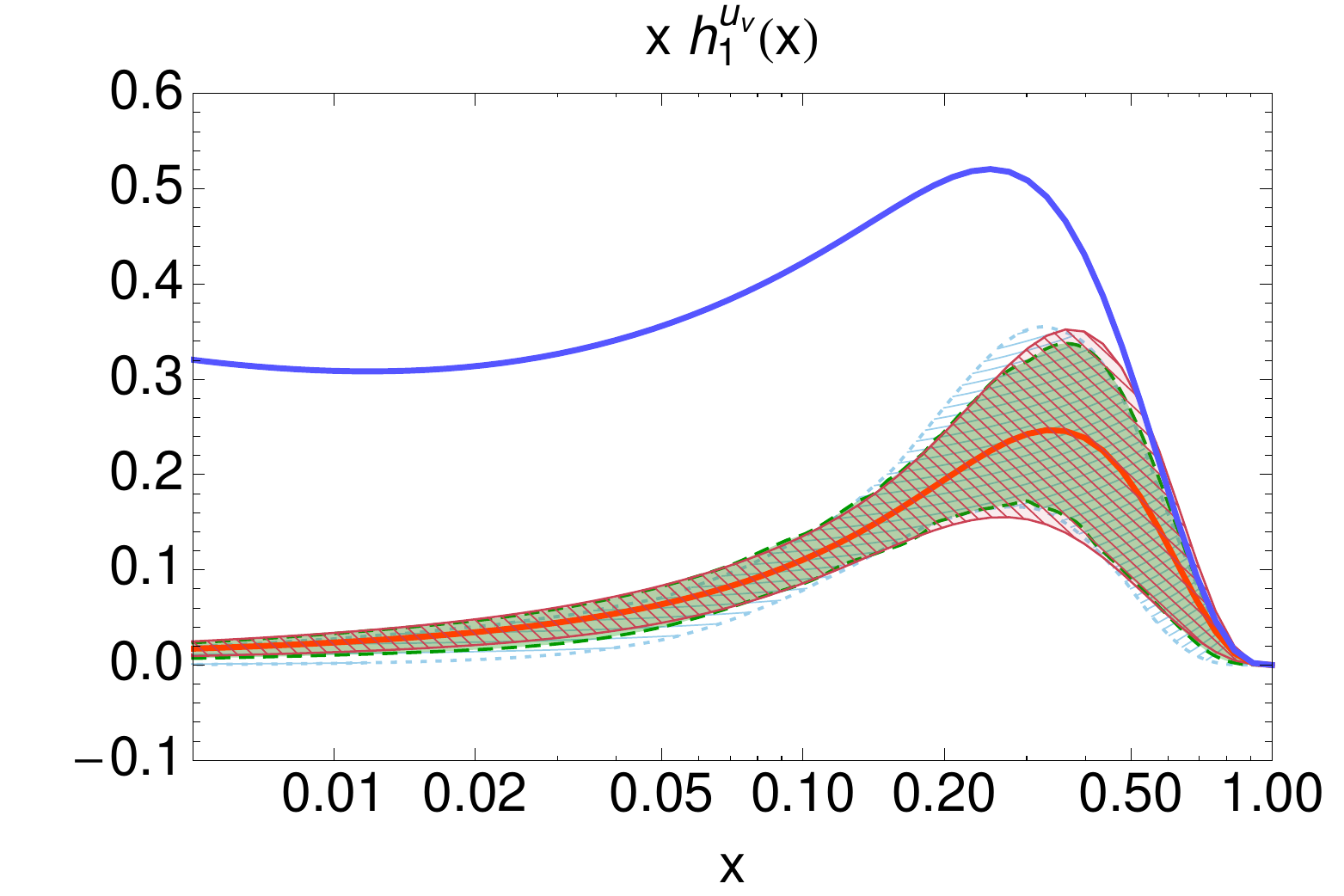}
   \includegraphics[width=0.49\textwidth]{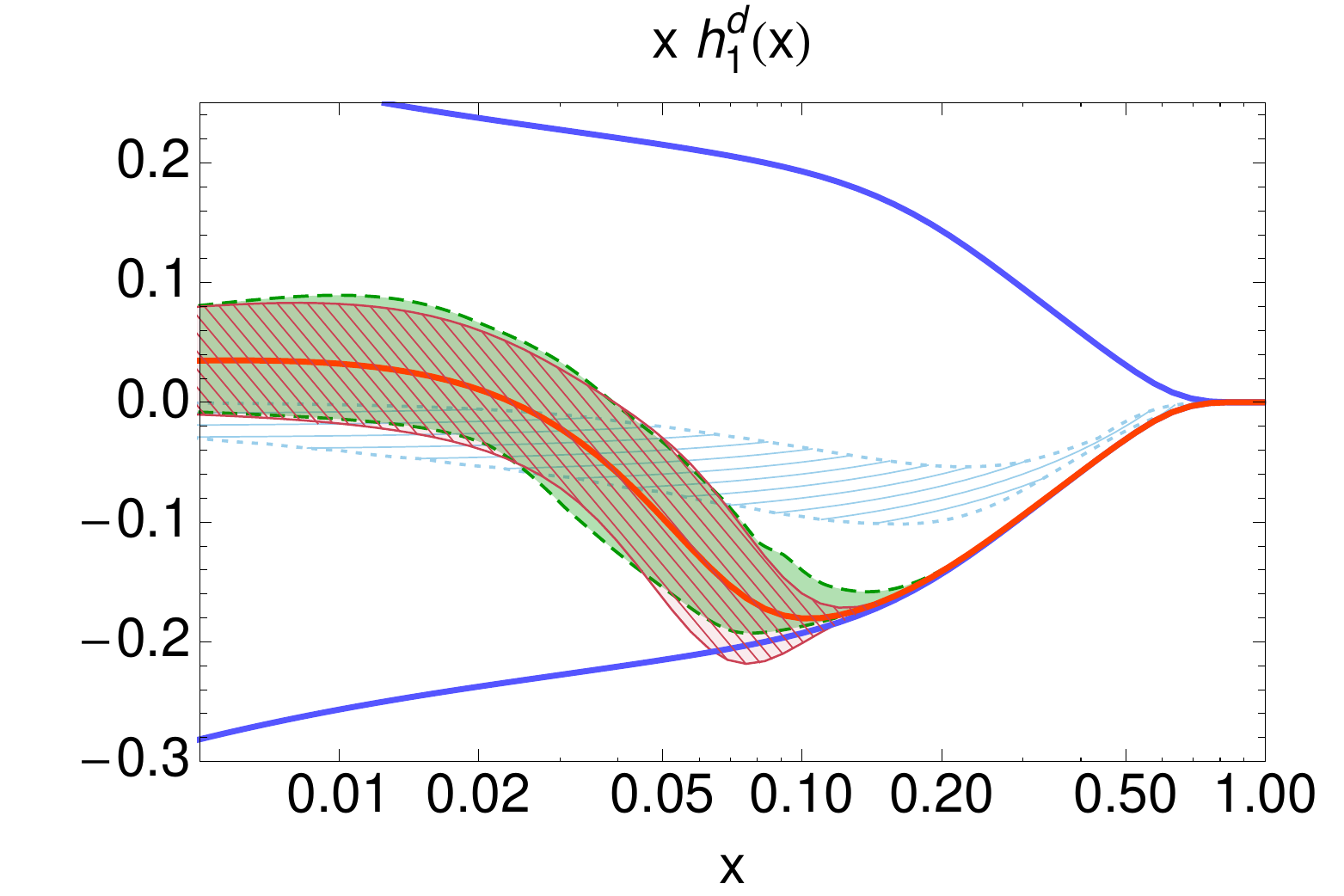}
\\
  \includegraphics[width=0.49\textwidth]{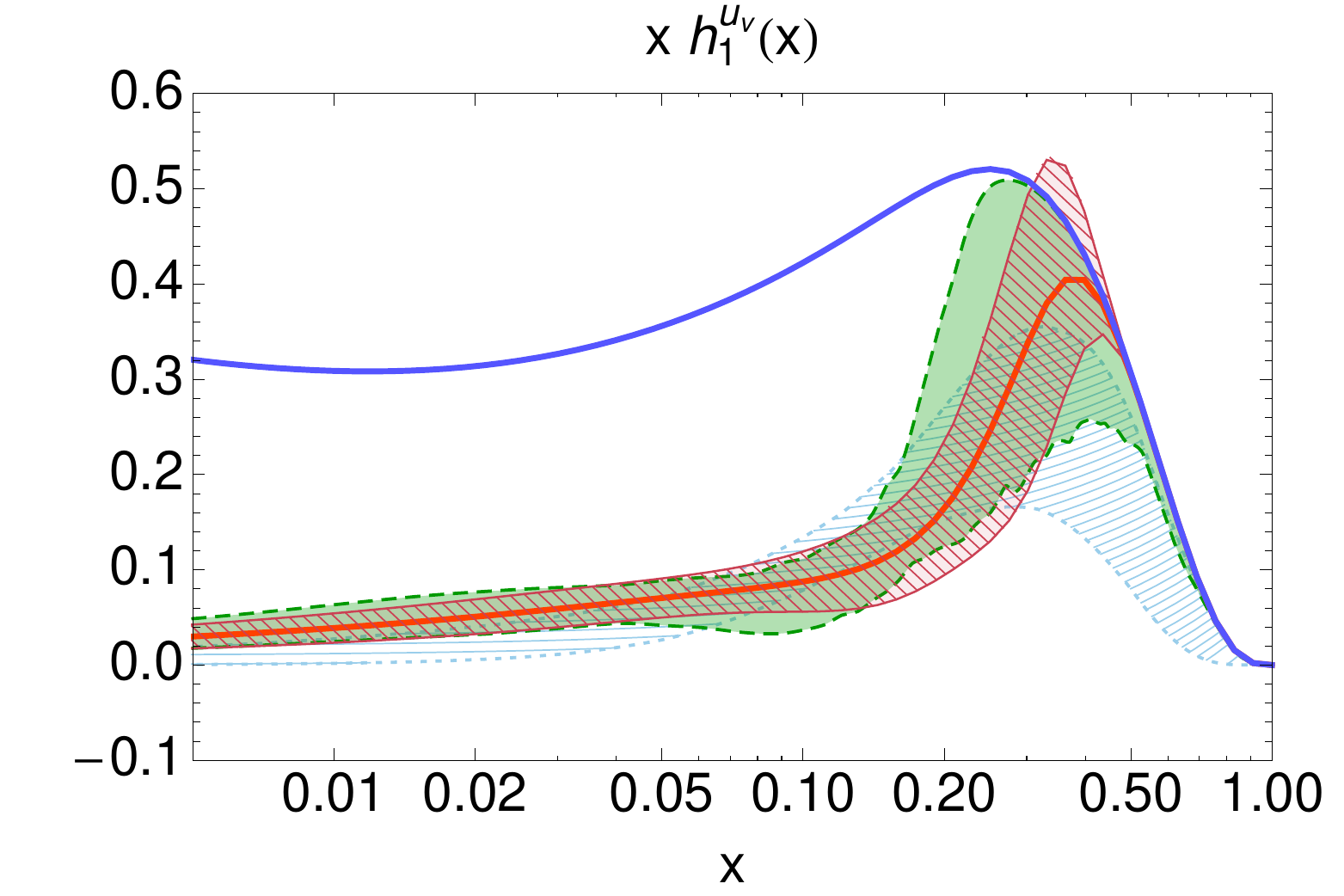}
   \includegraphics[width=0.49\textwidth]{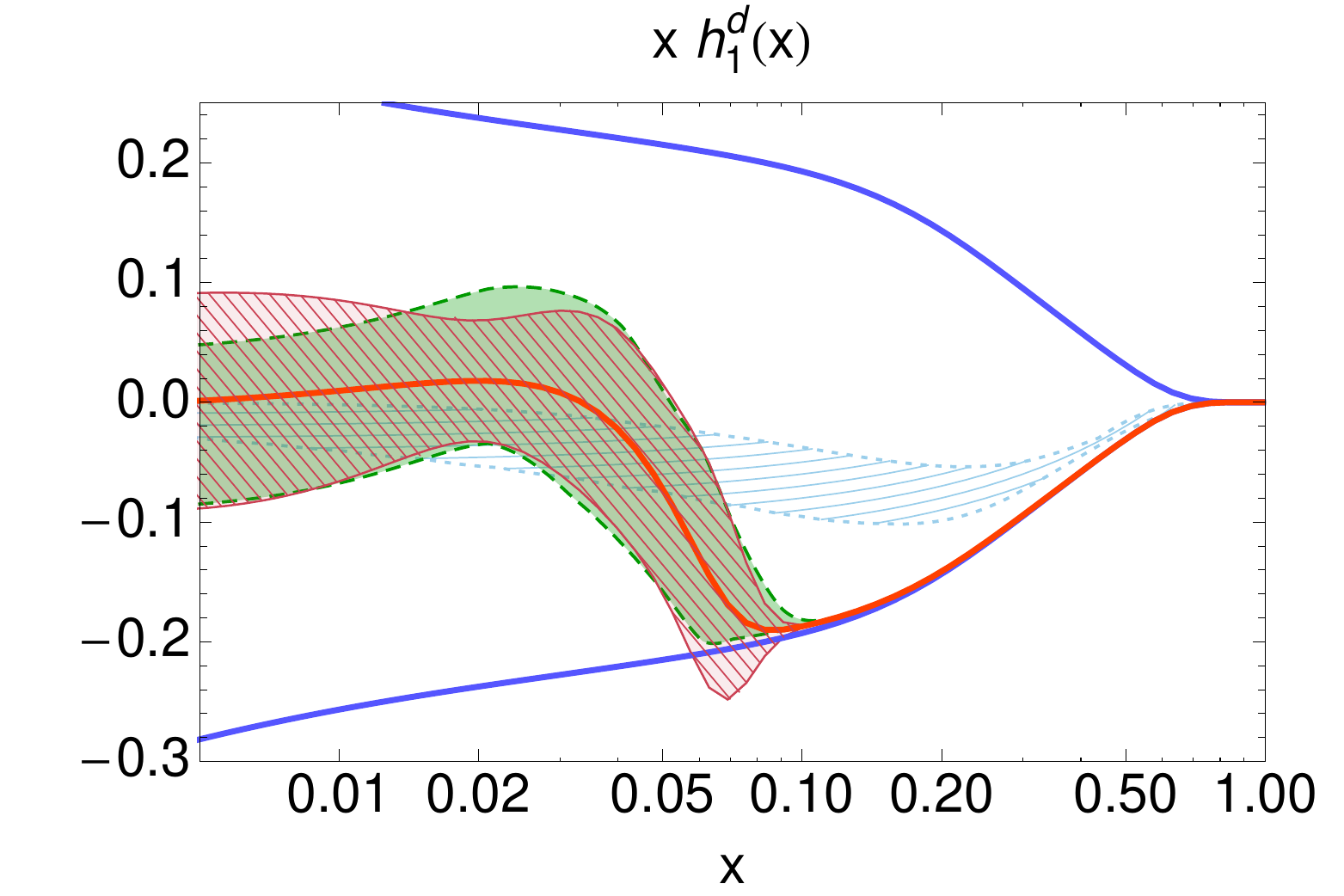}
\\
  \includegraphics[width=0.49\textwidth]{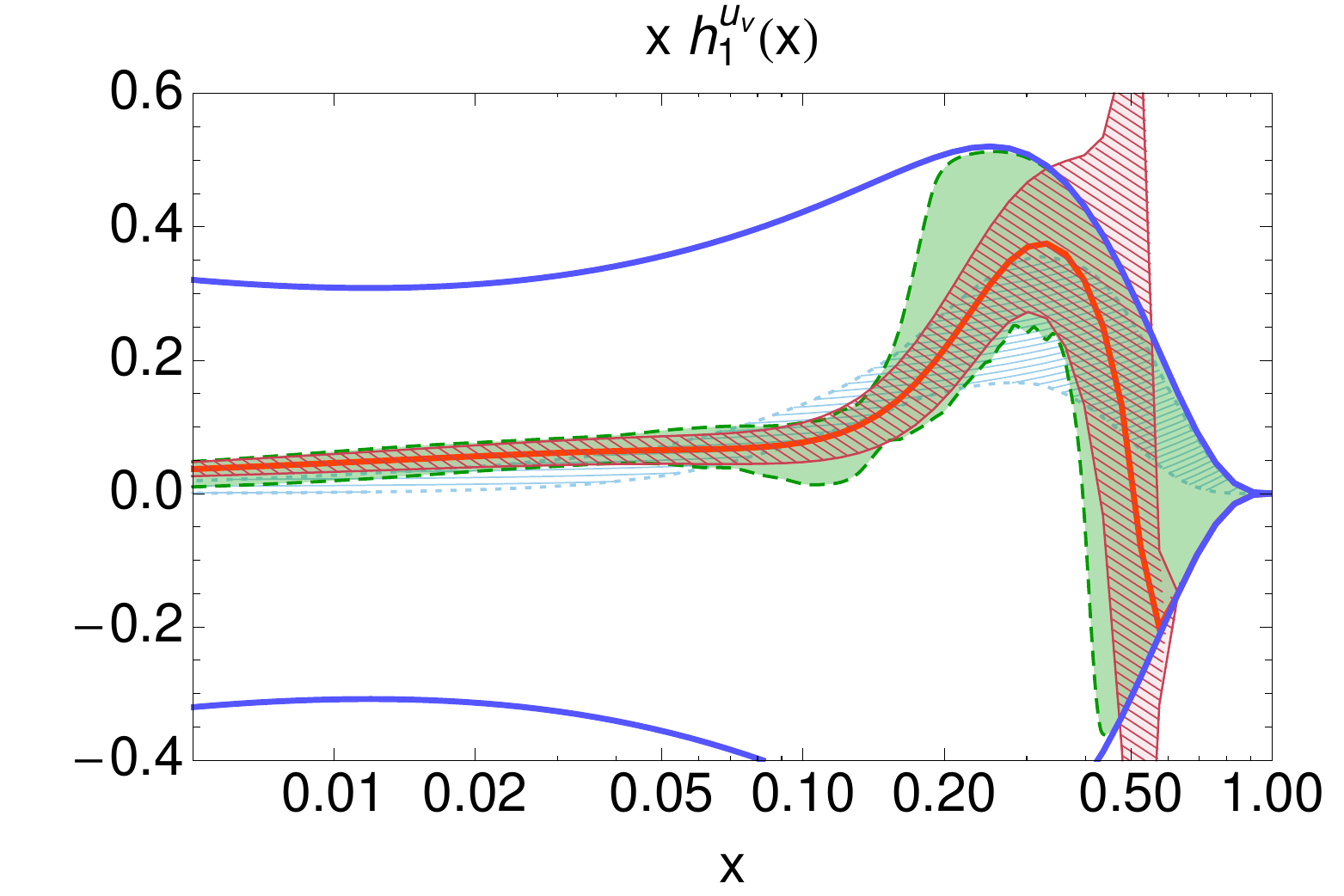}
   \includegraphics[width=0.49\textwidth]{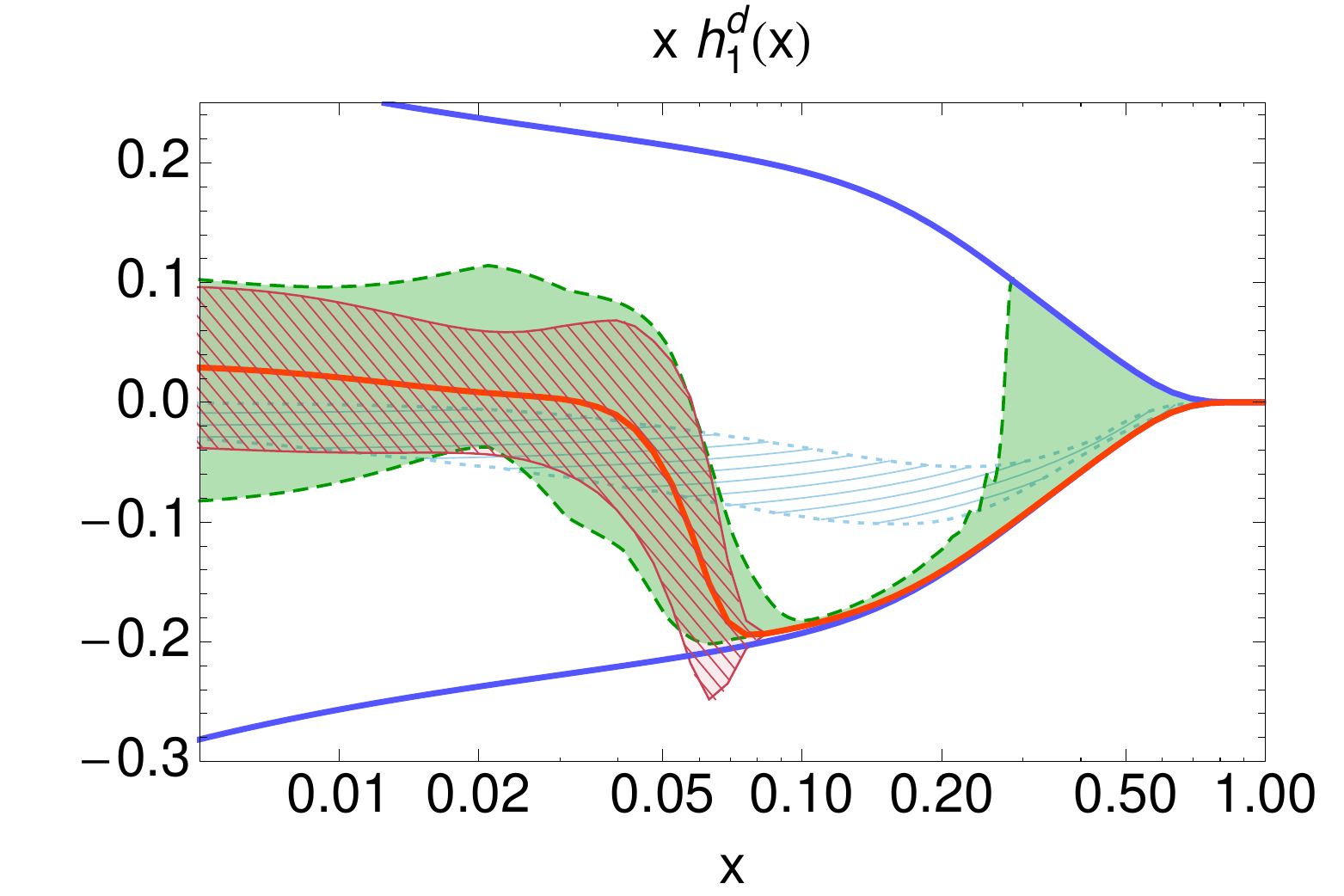}
  \caption{The up (left) and down (right) valence transversities coming from the present
    analysis evolved to $Q^2=2.4$ GeV$^2$. From top row to bottom, results with the 
    {\it rigid}, {\it flexible}, and {\it extra-flexible} scenarios are shown, respectively. 
    The dark thick solid lines are the Soffer bound. The uncertainty band with solid 
    boundaries is the best fit in the standard approach at $1\sigma$, whose central 
    value is given by the central thick solid line. The uncertainty band with dashed 
    boundaries is the $68\%$ of all fitting replicas obtained in the Monte Carlo approach. 
    As a comparison, the uncertainty band with short-dashed boundaries is the transversity 
    extraction from the Collins effect~\cite{Anselmino:2008jk}.}
  \label{f:xh1}
\end{figure}

The resulting transversity distribution is plotted in Fig.~\ref{f:xh1}. The left panel displays 
the $q = u_v$ contribution in Eq.~\eqref{e:funct_form}, while $q = d_v$ is in the right one. 
From top to bottom row, the results for the {\it rigid}, {\it flexible}, and {\it extra-flexible}, scenarios 
are shown, respectively. For each panel, the outcome in the standard approach with the 
Hessian method is represented by the uncertainty band with solid boundaries, the central 
thick solid line visualizing the central value. The partially overlapping band with dashed 
boundaries is the outcome when adopting the Monte Carlo approach, where the 
band width corresponds to the $68\%$ of all the 100 replicas, again produced as before by 
rejecting the largest $16\%$ and the lowest $16\%$ among the replicas' values in each $x$ point. 
As such, the set of selected replicas in the $68\%$ band can change in each different 
$x$ point; consequently, the band itself can show some irregular wiggles. For sake of 
comparison, each panel displays also the corresponding results for the only other 
existing parametrization available~\cite{Anselmino:2008jk}, depicted as a band with short-dashed 
boundaries. Since the latter was extracted at the scale $Q^2 = 2.4$ GeV$^2$, our 
results are properly evolved at the same scale. Finally, the dark thick solid lines 
indicate the Soffer bound, also evolved at the same scale $Q^2 = 2.4$ GeV$^2$ 
(using LO evolution as in the rest of the analysis). 

For the {\it flexible} scenario (middle row of Fig.~\ref{f:xh1}), the uncertainty bands 
in the standard and Monte Carlo approaches are quite similar. The main difference 
is that in the former case the boundaries of the band can occasionally cross the Soffer 
bound. This is due to the fact that the assumed quadratic dependence of $\chi^2$ 
on the parameters around its minimum is not a reliable one, when getting close to the bounds. 
On the contrary, in the Monte Carlo approach each replica is built such that it 
never violates the Soffer bound; the resulting $68\%$ band is always within 
those limits. 

For the valence up contribution (left panel), the standard approach tends
to saturate the Soffer bound at $x\sim 0.4$ (outside the range where data
exist). In the Monte Carlo approach, some of the replicas saturate the bound already 
at lower values. However, there are also a few replicas that do not saturate the bound at all, 
or even saturate the lower Soffer bound. These replicas typically fall outside the $68\%$ 
band drawn in the figures. Nevertheless, they can still have a good $\chi^2$ when 
compared to the data. 

For the down valence contribution (right panel), both 
approaches saturate the lower limit of the Soffer bound already at $x\sim 0.1$, i.e. in
a region where data exist. This behavior is driven by the data, in particular by the bins 
number 7 and 8 in the deuteron measurement. No such trend is evident in the 
corresponding single-hadron measurement of the Collins effect, from which the 
other parametrization of Ref.~\cite{Anselmino:2008jk} is extracted. As a matter of fact, this is the only 
source of significant discrepancy between the two extractions, which otherwise show 
a high level of compatibility despite the fact that they are obtained from very 
different procedures. Note that if the Soffer bound is saturated at some scale, 
it is likely to be significantly violated at a lower scale~\cite{Goldstein:1995ek}. 
Therefore, if we want to maintain the validity of
the Soffer bound at $Q^2 < 1$ GeV$^2$, we would expect
transversity to be clearly below the Soffer bound at $Q^2 \ge 1$ GeV$^2$. In fact, 
in our analysis with the Monte Carlo approach there are 
a few replicas that do not saturate the Soffer bound. They fall outside the $68\%$ 
band drawn in the figure, but they are still compatible with the data due to the large 
experimental error bars (this is true in particular for the deuteron bins number 7 and 8). 
Therefore, at present we cannot conclude that the the Soffer bound is saturated or violated,
even though the fit seems to point in that direction. We mention that
interesting speculations concerning violations of the Soffer bound were
presented in Ref.~\cite{Ralston:2008sm}. 

At low $x$, the functional form in the {\it flexible} scenario tends to zero by construction, 
and similarly in all other scenarios. However, the behavior down to $x\sim 0.005$ is 
driven by data. In fact, the functional form can have up to two nodes in $x \in [0,1]$. 
In the Monte Carlo approach, most of the replicas for $x h_1^{u_v}$ have no node, 
while for  $x h_1^{d_v}$ have one or even two nodes.

In the {\it rigid} scenario (upper row in Fig.~\ref{f:xh1}), most of the features are 
similar to the {\it flexible} scenario in the region where data exist, but there are 
some differences outside that range. For the valence up quark (left panel), both 
standard and Monte Carlo approaches give uncertainty bands that saturate the 
Soffer bound at higher $x$, almost completely overlapping with the result 
obtained from the Collins effect. For the valence down quark (right panel), the 
trend is very similar to the {\it flexible} scenario: the parametrization saturates 
the lower Soffer bound at $x \gtrsim 0.1$. This demonstrates that this unexpected 
behavior is not an artefact of the functional form, but is due to the experimental data.
In the Monte Carlo approach, the majority of replicas show the same behavior, but 
 a few ones (falling outside the 68\% band) do not, again as in the {\it flexible} scenario. 

Finally, for the {\it extra-flexible} scenario (bottom row in Fig.~\ref{f:xh1}) the 
distinction between regions with and without experimental data is even more clear. 
Where there are data, the results are highly compatible with the other scenarios. 
But at $x\gtrsim 0.4$ for the $u_v$ case (left panel), the uncertainty band in the 
standard approach substantially violates the Soffer bound. This is an artefact of  the 
assumptions used in error propagation, together with the lack of data at high $x$. 
In the Monte Carlo approach, the replicas entirely fill the area 
between the upper and lower Soffer bound, both for the up quark at $x\gtrsim 0.4$
and for the down quark at $x\gtrsim 0.25$. This is an explicit visualization of 
the realistic degree of uncertainty about transversity in the $x$ range where there are 
no experimental data points.

We have explored other scenarios for the transversity functional form. We
tried different arguments in the hyperbolic tangent, specifically different powers 
in the first factor, but with no significant change for $x\gtrsim 0.01$. Using 
the $x^{1/4}$ factor, the error band at low $x$ becomes considerably
wider. The valence transversities remain integrable, but stable values of the 
tensor charge can be reached only by pushing the lower limit in the integral 
to extremely small values $x\lesssim 10^{-10}$ (see below).

A qualitative comparison between the results of the present work and the 
available model predictions can be done using the results collected in 
Ref.~\cite{Bacchetta:2011bn}. In particular, we note that the extracted transversity 
for the up quark is smaller than most of the model calculations at intermediate 
$x \in [0.1,\  0.2]$, while it is larger at lower $x$ ($x\sim 0.01$). The down 
transversity is much larger in absolute value than all model calculations at 
intermediate $x$ (as observed before, this is due to the deuterium data points), 
while the error band is too large to draw any conclusion at lower $x$.


Transversity is directly related to the tensor charge, a fundamental quantity 
of hadrons at the same level as the vector, axial, and scalar charges. 
The tensor charge remains at the moment largely unconstrained. It 
can be directly compared with lattice QCD predictions (see, {\it e.g.},
Refs.~\cite{Aoki:1997pi,Gockeler:2006zu}) or models (see, {\it e.g.}, 
Refs.~\cite{Gamberg:2001qc,Wakamatsu:2007nc,Cloet:2007em,Lorce:2007fa,Ledwig:2010tu}). 
There is no sum rule related to the tensor current, due to the property of the
anomalous dimensions governing the QCD evolution of transversity. 
The contribution of a flavor $q$ to the tensor charge is defined as 
\begin{equation}
\delta q(Q^2)=\int \, dx\, h_1^{q_v}(x; Q^2) \; .
\label{e:tensor}
\end{equation}

\begin{table}
\begin{tabular}{|l|c|c|c|c|}
\hline
       &$\delta u$ & $\delta d$ &$\delta  u$ & $\delta d$ \\
\hline
$Q_0^2 = 1 \text{ GeV}^2$       &$x\in[0.0064,0.28]$ & $x\in[0.0064,0.28]$ & $x\in[0,1]$ & $x\in[0,1]$ 
\\
\hline
Standard rigid & $0.30\pm 0.09$ & $-0.26\pm 0.17$ & $0.57\pm 0.21$ & $-0.18 \pm 0.33$
 \\
\hline
MC rigid & $0.30\pm 0.07$ & $-0.22\pm 0.11$ & $0.56\pm 0.12$ & $-0.08 \pm 0.27$
 \\
\hline
Standard flex. & $0.29\pm 0.13$ & $-0.26\pm 0.22$ & $0.72\pm 0.24$ &$-0.33 \pm 0.61$
 \\
\hline
MC flex. & $0.32\pm 0.09$ & $-0.24\pm 0.11$ & $0.77\pm 0.22$ &$-0.45 \pm 0.48$
 \\
\hline
Standard extra-flex. & $0.32\pm 0.12$ & $-0.25\pm 0.15$ & $0.61\pm 0.40$ &$-0.16 \pm 0.44$
 \\
\hline
MC extra-flex. & $0.34\pm 0.10 $ & $-0.20\pm 0.14 $ & $0.68\pm 0.22$ &$-0.12 \pm 0.69$
 \\
\hline
\end{tabular}
\caption{Table of the results for the tensor charge at $Q_0^2 = 1$ GeV, truncated
  in the range where data exist (second and third column) and extended to the
  whole $x$ range (third and fourth column). The results are given for the
  standard and Monte Carlo approach and for the three scenarios considered in
  the fit.}
\label{t:tensor}
\end{table}

The region of validity of our fit is restricted to the experimental data
range. We can therefore give a reliable estimate for the tensor charge 
truncated to the interval $x \in [0.0064,0.28]$. In the first two columns from 
left of Tab.~\ref{t:tensor} we list the results obtained in the different approaches 
and scenarios. We tried also to extend the range of integration outside the experimental 
data to $x\in [0,1]$. The result is heavily influenced by the adopted functional
form, in particular by the low-$x$ exponent. Nevertheless, we quote our result
in the last two columns of Tab.~\ref{t:tensor}.

Our results for the tensor charges in the {\it flexible} scenario are slightly larger in absolute value 
compared to the ones in Ref.~\cite{Anselmino:2008jk}. They are compatible within errors.
Results obtained in several models can be found in, {\it e.g.},
Refs.~\cite{Barone:2001sp,Anselmino:2008jk,Wakamatsu:2008ki}. As emphasized in
Ref.~\cite{Wakamatsu:2008ki}, care must be taken when comparing results at
different scales. The ratio between the up and down tensor charges is scale
invariant. The values we obtain in all our scenarios are compatible with all 
models within the large errors. In order to better determine the tensor charge, more 
data at high and low $x$ are needed.


\section{Conclusions and Outlook}
\label{s:end}

The transversity parton distribution function (PDF) is 
an essential missing piece 
of our knowledge on the proton at leading-twist. 
It is a chiral-odd object, whose $Q^2$ 
dependence obeys the non-singlet QCD evolution. 
Its integral over $x$ is related to the nucleon tensor charge. 
Positivity bounds constrain its absolute
value to be smaller than the absolute value of the number density and helicity, the so-called 
Soffer bound. 
Due to its chiral-odd nature, transversity cannot be accessed in fully
  inclusive deep-inelastic scattering. It is however possible to access it in 
  two-particle-inclusive DIS~\cite{Collins:1994kq,Jaffe:1998hf,Radici:2001na}
  in combination with Dihadron Fragmentation Functions (DiFFs). 

In this paper, we have obtained for the first time the parameterization of the 
up and down valence transversities based on a collinear framework, 
using data for $\pi^+ \pi^-$ semi-inclusive DIS off transversely polarized targets 
from the HERMES and COMPASS collaborations~\cite{Airapetian:2008sk,Adolph:2012nw}, 
combined with the Belle data on almost back-to-back emission of two 
$\pi^+ \pi^-$ pairs in $e^+e^-$ annihilations~\cite{Vossen:2011fk}. We have explored different scenarios 
for the functional form, all subject to the theoretical constraint of the Soffer 
bound~\cite{Soffer:1995ww,Goldstein:1995ek}. We have also performed the error 
analysis in two independent ways. The first one is a standard one based on the 
Hessian method. The second one is based on the random generation of a large 
number of replicas of the experimental points, and on the fit of each of these 
replicas, producing an envelope of trajectories whose spread is the 
generalization of the $1\sigma$ uncertainty band when the distribution is not 
necessarily a Gaussian. 
As such, the second method is more reliable particularly when the fitting curves 
hit the Soffer bound, and the $\chi^2$ function cannot be expected to have 
the quadratic dependence on the fit parameters as required by the Hessian 
method. Nevertheless, in the kinematical range of the experimental 
measurements the two methods give almost overlapping results in all 
explored scenarios. 

In the range where data exist, 
our results are compatible with the only other existing parametrization of 
transversity, which is determined from the Collins effect in single-hadron 
SIDIS off transversely polarized targets~\cite{Anselmino:2008jk}. The only 
source of discrepancy lies in the range $0.1\lesssim x \lesssim 0.16$ for the
valence down quark, where two experimental data for the deuteron target 
drive our fitting curves to saturate the lower Soffer bound. However, the large 
error bars of these two points prevent us from drawing any conclusion about 
a possible violation of the Soffer inequality. 

Outside the kinematical range of experiments, the lack of data reflects itself
in a 
large uncertainty in the parametrization. For the {\it extra-flexible} scenario and 
the error analysis based on the random approach, the replicas take all 
the available values between the upper and lower Soffer bound at large $x$. 
This illustrates in a very effective way the need for new large$-x$ data in order to 
reduce the degree of uncertainty in the knowledge of transversity. 

In the near future, more data are expected from the HERMES and COMPASS 
collaborations. They will include also different types of hadron pairs (e.g., 
$K \pi$), which should allow us to improve the flavour separation of transversity.  
Two-particle inclusive DIS will be measured also at JLab in the future, 
which should considerably increase our knowledge of transversity at high $x$.   
Finally, invaluable information will come also from  
polarized proton-proton collisions~\cite{Bacchetta:2004it}: data are expected 
from the PHENIX and STAR collaborations (see, e.g.,~\cite{Yang:2009zzr}).


\section*{Acknowledgments}
We are grateful to Marco Guagnelli for providing us with a modified version of the HOPPET evolution code.
We acknowledge useful discussions with Andrea Bianconi, Stefano Melis, and Emanuele Nocera.
 A.~Courtoy is working under the Belgian Fund F.R.S.-FNRS via the contract of
 Charg\'ee de recherches. This work is partially supported by the Italian MIUR
 through the PRIN 2008EKLACK, and by the Joint Research Activity ``Study of
 Strongly Interacting Matter" (acronym HadronPhysics3, Grant Agreement
 No. 283286) under the 7th Framework Programme of the European Community.

 \appendix
 \section{Appendix}
\label{s:app}

For convenience, we reproduce here the explicit analytic forms of the Soffer
bound used in our analysis, which is based on the MSTW08LO set~\cite{Martin:2009iq} for the unpolarized PDF, combined to the DSSV parameterization~\cite{deFlorian:2009vb} for the helicity distribution. The
equations hold at $Q_0^2 = 1$ GeV$^2$.
\begin{align}
x \mbox{\small SB}^{u}(x)+x \mbox{\small SB}^{\bar{u}}(x) & = 
\frac{1}{2}\Bigl[x f_1^{u_v}(x) + 2 x f_1^{\bar{u}}(x) 
+ x g_1^{u}(x)+x g_1^{\bar{u}}(x)\Bigr],
\\
x f_1^{u_v}(x)&= 1.4335  x^{0.45232} (1-x)^{3.0409}
\left(1+8.9924 x-2.3737 \sqrt{x}\right),
\\
\begin{split} 
2 x f_1^{\bar{u}}(x)&=
\frac{1}{2} \Bigl[\left(1+16.865 x-2.9012 \sqrt{x}\right)
\\&\quad
\times \frac{0.59964  (1-x)^{8.8801}-0.10302   (1-x)^{13.242}}{x^{0.16276}}
\\ & \quad
-17.8826 x^{1.876}(1-x)^{10.8801}
\left(1-36.507 x^2+8.4703 x\right) 
\Bigr],
\end{split} 
\\
x g_1^{u}(x)+x g_1^{\bar{u}}(x) &= 0.677 x^{0.692} (1-x)^{3.34} \left(1+15.87 x-2.18 \sqrt{x}\right) ,
\\
\nonumber
\\
x \mbox{\small SB}^{d}(x)+x \mbox{\small SB}^{\bar{d}}(x) & = 
\frac{1}{2}\Bigl[x f_1^{d_v}(x) + 2 x f_1^{\bar{d}}(x) 
+ x g_1^{d}(x)+x g_1^{\bar{d}}(x)\Bigr],
\\
x f_1^{d_v}(x)&= 
5.0903  x^{0.71978} (1-x)^{5.1244}\left(1+7.473 x-4.3654 \sqrt{x}\right),
\\
\begin{split} 
2 x f_1^{\bar{d}}(x)&= 2 x f_1^{\bar{u}}(x)
\\ & \quad
+17.8826 x^{1.876} (1-x)^{10.8801} \left(1-36.507 x^2+8.4703 x\right)
,
\end{split} 
\\
x g_1^{d}(x)+x g_1^{\bar{d}}(x) &= -0.015 x^{0.164} (1-x)^{3.89} 
\left(1+98.94 x+22.4 \sqrt{x}\right)
.
\end{align}


\bibliographystyle{jhep}
\bibliography{mybiblio}

\providecommand{\href}[2]{#2}\begingroup\raggedright\begin{thebibliography}{10}

\bibitem{Ralston:1979ys}
J.~P. Ralston and D.~E. Soper, {\it Production of dimuons from high-energy
  polarized proton- proton collisions},  {\em Nucl. Phys.} {\bf B152} (1979)
  109.

\bibitem{Artru:1990zv}
X.~Artru and M.~Mekhfi, {\it Transversely polarized parton densities, their
  evolution and their measurement},  {\em Z. Phys.} {\bf C45} (1990) 669--676.

\bibitem{Jaffe:1991kp}
R.~L. Jaffe and X.~Ji, {\it Chiral odd parton distributions and polarized
  drell-yan},  {\em Phys. Rev. Lett.} {\bf 67} (1991) 552--555.

\bibitem{Cortes:1991ja}
J.~Cortes, B.~Pire, and J.~Ralston, {\it {Measuring the transverse polarization
  of quarks in the proton}},  {\em Z.Phys.} {\bf C55} (1992) 409--416.

\bibitem{Blumlein:2012bf}
J.~Blumlein, {\it {The Theory of Deeply Inelastic Scattering}},
  \href{http://xxx.lanl.gov/abs/1208.6087}{{\tt arXiv:1208.6087}}.

\bibitem{Leader:2010rb}
E.~Leader, A.~V. Sidorov, and D.~B. Stamenov, {\it {Determination of Polarized
  PDFs from a QCD Analysis of Inclusive and Semi-inclusive Deep Inelastic
  Scattering Data}},  {\em Phys.Rev.} {\bf D82} (2010) 114018,
  [\href{http://xxx.lanl.gov/abs/1010.0574}{{\tt arXiv:1010.0574}}].

\bibitem{deFlorian:2009vb}
D.~de~Florian, R.~Sassot, M.~Stratmann, and W.~Vogelsang, {\it {Extraction of
  Spin-Dependent Parton Densities and Their Uncertainties}},  {\em Phys.Rev.}
  {\bf D80} (2009) 034030.

\bibitem{Hirai:2003pm}
{\bf Asymmetry Analysis} Collaboration, M.~Hirai, S.~Kumano, and N.~Saito, {\it
  {Determination of polarized parton distribution functions and their
  uncertainties}},  {\em Phys. Rev.} {\bf D69} (2004) 054021,
  [\href{http://xxx.lanl.gov/abs/hep-ph/0312112}{{\tt hep-ph/0312112}}].

\bibitem{Bluemlein:2002be}
J.~Blumlein and H.~Bottcher, {\it {QCD analysis of polarized deep inelastic
  data and parton distributions}},  {\em Nucl.Phys.} {\bf B636} (2002)
  225--263, [\href{http://xxx.lanl.gov/abs/hep-ph/0203155}{{\tt
  hep-ph/0203155}}].

\bibitem{D'Alesio:2012pp}
U.~D'Alesio, {\it {Transversity}},
  \href{http://xxx.lanl.gov/abs/1209.4773}{{\tt arXiv:1209.4773}}.

\bibitem{Jaffe:1997yz}
R.~L. Jaffe, {\it Can transversity be measured?},
  \href{http://xxx.lanl.gov/abs/http://arXiv.org/abs/hep-ph/9710465}{{\tt
  http://arXiv.org/abs/hep-ph/9710465}}.

\bibitem{Airapetian:2004tw}
{\bf HERMES} Collaboration, A.~Airapetian et~al., {\it Single-spin asymmetries
  in semi-inclusive deep-inelastic scattering on a transversely polarized
  hydrogen target},  {\em Phys. Rev. Lett.} {\bf 94} (2005) 012002,
  [\href{http://xxx.lanl.gov/abs/hep-ex/0408013}{{\tt hep-ex/0408013}}].

\bibitem{Ageev:2006da}
{\bf COMPASS} Collaboration, E.~S. Ageev et~al., {\it A new measurement of the
  collins and sivers asymmetries on a transversely polarised deuteron target},
  {\em Nucl. Phys.} {\bf B765} (2007) 31--70,
  [\href{http://xxx.lanl.gov/abs/hep-ex/0610068}{{\tt hep-ex/0610068}}].

\bibitem{Abe:2005zx}
{\bf Belle} Collaboration, K.~Abe et~al., {\it {Measurement of azimuthal
  asymmetries in inclusive production of hadron pairs in e+ e- annihilation at
  Belle}},  {\em Phys.Rev.Lett.} {\bf 96} (2006) 232002,
  [\href{http://xxx.lanl.gov/abs/hep-ex/0507063}{{\tt hep-ex/0507063}}].

\bibitem{Anselmino:2008jk}
M.~Anselmino, M.~Boglione, U.~D'Alesio, A.~Kotzinian, F.~Murgia, A.~Prokudin,
  and S.~Melis, {\it {Update on transversity and Collins functions from SIDIS
  and e+ e- data}},  {\em Nucl. Phys. Proc. Suppl.} {\bf 191} (2009) 98--107,
  [\href{http://xxx.lanl.gov/abs/0812.4366}{{\tt arXiv:0812.4366}}].

\bibitem{Collins:2011zzd}
J.~Collins, {\em Foundations of Perturbative QCD}.
\newblock Cambridge Monographs on Particle Physics, Nuclear Physics and
  Cosmology. Cambridge University Press, 2011.

\bibitem{Aybat:2011zv}
S.~Aybat and T.~C. Rogers, {\it {TMD Parton Distribution and Fragmentation
  Functions with QCD Evolution}},  {\em Phys. Rev.} {\bf D83} (2011) 114042,
  [\href{http://xxx.lanl.gov/abs/1101.5057}{{\tt arXiv:1101.5057}}].

\bibitem{Aybat:2011ge}
S.~M. Aybat, J.~C. Collins, J.-W. Qiu, and T.~C. Rogers, {\it {The QCD
  Evolution of the Sivers Function}},  {\em Phys.Rev.} {\bf D85} (2012) 034043,
  [\href{http://xxx.lanl.gov/abs/1110.6428}{{\tt arXiv:1110.6428}}].

\bibitem{Echevarria:2012pw}
M.~G. Echevarria, A.~Idilbi, A.~Schafer, and I.~Scimemi, {\it
  {Model-Independent Evolution of Transverse Momentum Dependent Distribution
  Functions (TMDs) at NNLL}},  \href{http://xxx.lanl.gov/abs/1208.1281}{{\tt
  arXiv:1208.1281}}.

\bibitem{Prokudin:2012fj}
A.~Prokudin, {\it {QCD Evolution Workshop: Introduction}},
  \href{http://xxx.lanl.gov/abs/1210.4133}{{\tt arXiv:1210.4133}}.

\bibitem{Anselmino:2012re}
M.~Anselmino, M.~Boglione, and S.~Melis, {\it {Phenomenology of Sivers Effect
  with TMD Evolution}},  \href{http://xxx.lanl.gov/abs/1209.1541}{{\tt
  arXiv:1209.1541}}.

\bibitem{Collins:1994kq}
J.~C. Collins, S.~F. Heppelmann, and G.~A. Ladinsky, {\it Measuring
  transversity densities in singly polarized hadron-hadron and lepton-hadron
  collisions},  {\em Nucl. Phys.} {\bf B420} (1994) 565--582,
  [\href{http://xxx.lanl.gov/abs/http://arXiv.org/abs/hep-ph/9305309}{{\tt
  http://arXiv.org/abs/hep-ph/9305309}}].

\bibitem{Jaffe:1998hf}
R.~L. Jaffe, X.~Jin, and J.~Tang, {\it Interference fragmentation functions and
  the nucleon's transversity},  {\em Phys. Rev. Lett.} {\bf 80} (1998)
  1166--1169,
  [\href{http://xxx.lanl.gov/abs/http://arXiv.org/abs/hep-ph/9709322}{{\tt
  http://arXiv.org/abs/hep-ph/9709322}}].

\bibitem{Radici:2001na}
M.~Radici, R.~Jakob, and A.~Bianconi, {\it Accessing transversity with
  interference fragmentation functions},  {\em Phys. Rev.} {\bf D65} (2002)
  074031,
  [\href{http://xxx.lanl.gov/abs/http://arXiv.org/abs/hep-ph/0110252}{{\tt
  http://arXiv.org/abs/hep-ph/0110252}}].

\bibitem{Ceccopieri:2007ip}
F.~A. Ceccopieri, M.~Radici, and A.~Bacchetta, {\it {Evolution equations for
  extended dihadron fragmentation functions}},  {\em Phys.Lett.} {\bf B650}
  (2007) 81--89, [\href{http://xxx.lanl.gov/abs/hep-ph/0703265}{{\tt
  hep-ph/0703265}}].

\bibitem{Airapetian:2008sk}
{\bf HERMES} Collaboration, A.~Airapetian et~al., {\it {Evidence for a
  Transverse Single-Spin Asymmetry in Leptoproduction of pi+pi- Pairs}},  {\em
  JHEP} {\bf 06} (2008) 017, [\href{http://xxx.lanl.gov/abs/0803.2367}{{\tt
  arXiv:0803.2367}}].

\bibitem{Bacchetta:2002ux}
A.~Bacchetta and M.~Radici, {\it Partial-wave analysis of two-hadron
  fragmentation functions},  {\em Phys. Rev.} {\bf D67} (2003) 094002,
  [\href{http://xxx.lanl.gov/abs/hep-ph/0212300}{{\tt hep-ph/0212300}}].

\bibitem{Diehl:2005pc}
M.~Diehl and S.~Sapeta, {\it On the analysis of lepton scattering on
  longitudinally or transversely polarized protons},  {\em Eur. Phys. J.} {\bf
  C41} (2005) 515--533, [\href{http://xxx.lanl.gov/abs/hep-ph/0503023}{{\tt
  hep-ph/0503023}}].

\bibitem{Bacchetta:2006tn}
A.~Bacchetta, M.~Diehl, K.~Goeke, A.~Metz, P.~J. Mulders, and M.~Schlegel, {\it
  Semi-inclusive deep inelastic scattering at small transverse momentum},  {\em
  JHEP} {\bf 02} (2007) 093,
  [\href{http://xxx.lanl.gov/abs/hep-ph/0611265}{{\tt hep-ph/0611265}}].

\bibitem{Bianconi:1999cd}
A.~Bianconi, S.~Boffi, R.~Jakob, and M.~Radici, {\it Two-hadron interference
  fragmentation functions. i: General framework},  {\em Phys. Rev.} {\bf D62}
  (2000) 034008,
  [\href{http://xxx.lanl.gov/abs/http://arXiv.org/abs/hep-ph/9907475}{{\tt
  http://arXiv.org/abs/hep-ph/9907475}}].

\bibitem{Zhou:2011ba}
J.~Zhou and A.~Metz, {\it {Dihadron fragmentation functions for large invariant
  mass}},  {\em Phys.Rev.Lett.} {\bf 106} (2011) 172001,
  [\href{http://xxx.lanl.gov/abs/1101.3273}{{\tt arXiv:1101.3273}}].

\bibitem{Bacchetta:2006un}
A.~Bacchetta and M.~Radici, {\it Modeling dihadron fragmentation functions},
  {\em Phys. Rev.} {\bf D74} (2006) 114007,
  [\href{http://xxx.lanl.gov/abs/hep-ph/0608037}{{\tt hep-ph/0608037}}].

\bibitem{Bacchetta:2011ip}
A.~Bacchetta, A.~Courtoy, and M.~Radici, {\it {First glances at the
  transversity parton distribution through dihadron fragmentation functions}},
  {\em Phys.Rev.Lett.} {\bf 107} (2011) 012001,
  [\href{http://xxx.lanl.gov/abs/1104.3855}{{\tt arXiv:1104.3855}}].

\bibitem{Vossen:2011fk}
{\bf Belle Collaboration} Collaboration, A.~Vossen et~al., {\it {Observation of
  transverse polarization asymmetries of charged pion pairs in e+e-
  annihilation near sqrt s=10.58 GeV}},  {\em Phys.Rev.Lett.} {\bf 107} (2011)
  072004, [\href{http://xxx.lanl.gov/abs/1104.2425}{{\tt arXiv:1104.2425}}].

\bibitem{Bianconi:1999uc}
A.~Bianconi, S.~Boffi, R.~Jakob, and M.~Radici, {\it Two-hadron interference
  fragmentation functions. ii: A model calculation},  {\em Phys. Rev.} {\bf
  D62} (2000) 034009,
  [\href{http://xxx.lanl.gov/abs/http://arXiv.org/abs/hep-ph/9907488}{{\tt
  http://arXiv.org/abs/hep-ph/9907488}}].

\bibitem{Casey:2012ux}
A.~Casey, H.~H. Matevosyan, and A.~W. Thomas, {\it {Calculating Dihadron
  Fragmentation Functions in the NJL-jet model}},  {\em Phys.Rev.} {\bf D85}
  (2012) 114049, [\href{http://xxx.lanl.gov/abs/1202.4036}{{\tt
  arXiv:1202.4036}}].

\bibitem{Bacchetta:2008wb}
A.~Bacchetta, F.~A. Ceccopieri, A.~Mukherjee, and M.~Radici, {\it {Asymmetries
  involving dihadron fragmentation functions: from DIS to e+e- annihilation}},
  {\em Phys.Rev.} {\bf D79} (2009) 034029,
  [\href{http://xxx.lanl.gov/abs/0812.0611}{{\tt arXiv:0812.0611}}].

\bibitem{Boer:2003ya}
D.~Boer, R.~Jakob, and M.~Radici, {\it Interference fragmentation functions in
  electron positron annihilation},  {\em Phys. Rev.} {\bf D67} (2003) 094003,
  [\href{http://xxx.lanl.gov/abs/hep-ph/0302232}{{\tt hep-ph/0302232}}].

\bibitem{Courtoy:2012ry}
A.~Courtoy, A.~Bacchetta, M.~Radici, and A.~Bianconi, {\it {First extraction of
  Interference Fragmentation Functions from $e^+e^-$ data}},  {\em Phys.Rev.}
  {\bf D85} (2012) 114023, [\href{http://xxx.lanl.gov/abs/1202.0323}{{\tt
  arXiv:1202.0323}}].

\bibitem{Salam:2008qg}
G.~P. Salam and J.~Rojo, {\it {A Higher Order Perturbative Parton Evolution
  Toolkit (HOPPET)}},  {\em Comput.Phys.Commun.} {\bf 180} (2009) 120--156,
  [\href{http://xxx.lanl.gov/abs/0804.3755}{{\tt arXiv:0804.3755}}].

\bibitem{Adolph:2012nw}
{\bf COMPASS Collaboration} Collaboration, C.~Adolph et~al., {\it {Transverse
  spin effects in hadron-pair production from semi-inclusive deep inelastic
  scattering}},  {\em Phys.Lett.} {\bf B713} (2012) 10--16,
  [\href{http://xxx.lanl.gov/abs/1202.6150}{{\tt arXiv:1202.6150}}].

\bibitem{Elia:2012}
C.~Elia, {\em Measurement of two-hadron transverse spinasymmetries in SIDIS at
  COMPASS}.
\newblock PhD thesis, Trieste U., 2012.
\newblock {http://hdl.handle.net/10077/7425}.

\bibitem{Martin:2009iq}
A.~D. Martin, W.~J. Stirling, R.~S. Thorne, and G.~Watt, {\it {Parton
  distributions for the LHC}},  {\em Eur. Phys. J.} {\bf C63} (2009) 189--285.

\bibitem{Courtoy:2012in}
A.~Courtoy, A.~Bacchetta, and M.~Radici, {\it {Status on the transversity
  parton distribution: the dihadron fragmentation functions way}},  {\em PoS}
  {\bf QNP2012} (2012) 042, [\href{http://xxx.lanl.gov/abs/1206.1836}{{\tt
  arXiv:1206.1836}}].

\bibitem{Soffer:1995ww}
J.~Soffer, {\it Positivity constraints for spin dependent parton
  distributions},  {\em Phys. Rev. Lett.} {\bf 74} (1995) 1292--1294,
  [\href{http://xxx.lanl.gov/abs/http://arXiv.org/abs/hep-ph/9409254}{{\tt
  http://arXiv.org/abs/hep-ph/9409254}}].

\bibitem{Goldstein:1995ek}
G.~R. Goldstein, R.~Jaffe, and X.-D. Ji, {\it {Soffer's inequality}},  {\em
  Phys.Rev.} {\bf D52} (1995) 5006--5013,
  [\href{http://xxx.lanl.gov/abs/hep-ph/9501297}{{\tt hep-ph/9501297}}].

\bibitem{Bourrely:1998bx}
C.~Bourrely, J.~Soffer, and O.~V. Teryaev, {\it The q**2 evolution of soffer
  inequality},  {\em Phys. Lett.} {\bf B420} (1998) 375--381,
  [\href{http://xxx.lanl.gov/abs/http://arXiv.org/abs/hep-ph/9710224}{{\tt
  http://arXiv.org/abs/hep-ph/9710224}}].

\bibitem{Vogelsang:1997ak}
W.~Vogelsang, {\it {Next-to-leading order evolution of transversity
  distributions and Soffer's inequality}},  {\em Phys.Rev.} {\bf D57} (1998)
  1886--1894, [\href{http://xxx.lanl.gov/abs/hep-ph/9706511}{{\tt
  hep-ph/9706511}}].

\bibitem{Forte:2002fg}
S.~Forte, L.~Garrido, J.~I. Latorre, and A.~Piccione, {\it {Neural network
  parametrization of deep inelastic structure functions}},  {\em JHEP} {\bf
  0205} (2002) 062, [\href{http://xxx.lanl.gov/abs/hep-ph/0204232}{{\tt
  hep-ph/0204232}}].

\bibitem{Ball:2008by}
{\bf NNPDF Collaboration} Collaboration, R.~D. Ball et~al., {\it {A
  Determination of parton distributions with faithful uncertainty estimation}},
   {\em Nucl.Phys.} {\bf B809} (2009) 1--63,
  [\href{http://xxx.lanl.gov/abs/0808.1231}{{\tt arXiv:0808.1231}}].

\bibitem{Ball:2010de}
R.~D. Ball, L.~Del~Debbio, S.~Forte, A.~Guffanti, J.~I. Latorre, et~al., {\it
  {A first unbiased global NLO determination of parton distributions and their
  uncertainties}},  {\em Nucl.Phys.} {\bf B838} (2010) 136--206.

\bibitem{Ralston:2008sm}
J.~P. Ralston, {\it {Exploring Confinement with Spin}},
  \href{http://xxx.lanl.gov/abs/0810.0871}{{\tt arXiv:0810.0871}}.

\bibitem{Bacchetta:2011bn}
A.~Bacchetta, {\it {Models for Transverse-Momentum Distributions and
  Transversity}},  {\em Nuovo Cim.} {\bf C035N2} (2012) 19--28,
  [\href{http://xxx.lanl.gov/abs/1111.6642}{{\tt arXiv:1111.6642}}].

\bibitem{Aoki:1997pi}
S.~Aoki, M.~Doui, T.~Hatsuda, and Y.~Kuramashi, {\it Tensor charge of the
  nucleon in lattice qcd},  {\em Phys. Rev.} {\bf D56} (1997) 433--436,
  [\href{http://xxx.lanl.gov/abs/http://arXiv.org/abs/hep-lat/9608115}{{\tt
  http://arXiv.org/abs/hep-lat/9608115}}].

\bibitem{Gockeler:2006zu}
{\bf QCDSF} Collaboration, M.~Gockeler et~al., {\it {Transverse spin structure
  of the nucleon from lattice QCD simulations}},  {\em Phys. Rev. Lett.} {\bf
  98} (2007) 222001, [\href{http://xxx.lanl.gov/abs/hep-lat/0612032}{{\tt
  hep-lat/0612032}}].

\bibitem{Gamberg:2001qc}
L.~P. Gamberg and G.~R. Goldstein, {\it {Flavor spin symmetry estimate of the
  nucleon tensor charge}},  {\em Phys.Rev.Lett.} {\bf 87} (2001) 242001,
  [\href{http://xxx.lanl.gov/abs/hep-ph/0107176}{{\tt hep-ph/0107176}}].

\bibitem{Wakamatsu:2007nc}
M.~Wakamatsu, {\it {Comparative analysis of the transversities and the
  longitudinally polarized distribution functions of the nucleon}},  {\em Phys.
  Lett.} {\bf B653} (2007) 398--403,
  [\href{http://xxx.lanl.gov/abs/0705.2917}{{\tt arXiv:0705.2917}}].

\bibitem{Cloet:2007em}
I.~C. Cloet, W.~Bentz, and A.~W. Thomas, {\it {Transversity quark distributions
  in a covariant quark- diquark model}},  {\em Phys. Lett.} {\bf B659} (2008)
  214--220, [\href{http://xxx.lanl.gov/abs/0708.3246}{{\tt arXiv:0708.3246}}].

\bibitem{Lorce:2007fa}
C.~Lorce, {\it {Tensor charges of light baryons in the Infinite Momentum
  Frame}},  {\em Phys.Rev.} {\bf D79} (2009) 074027,
  [\href{http://xxx.lanl.gov/abs/0708.4168}{{\tt arXiv:0708.4168}}].

\bibitem{Ledwig:2010tu}
T.~Ledwig, A.~Silva, and H.-C. Kim, {\it {Tensor charges and form factors of
  SU(3) baryons in the self-consistent SU(3) chiral quark-soliton model}},
  {\em Phys.Rev.} {\bf D82} (2010) 034022,
  [\href{http://xxx.lanl.gov/abs/1004.3612}{{\tt arXiv:1004.3612}}].

\bibitem{Barone:2001sp}
V.~Barone, A.~Drago, and P.~G. Ratcliffe, {\it Transverse polarisation of
  quarks in hadrons},  {\em Phys. Rept.} {\bf 359} (2002) 1--168,
  [\href{http://xxx.lanl.gov/abs/http://arXiv.org/abs/hep-ph/0104283}{{\tt
  http://arXiv.org/abs/hep-ph/0104283}}].

\bibitem{Wakamatsu:2008ki}
M.~Wakamatsu, {\it {Chiral-odd GPDs, transversity decomposition of angular
  momentum, and tensor charges of the nucleon}},  {\em Phys.Rev.} {\bf D79}
  (2009) 014033, [\href{http://xxx.lanl.gov/abs/0811.4196}{{\tt
  arXiv:0811.4196}}].

\bibitem{Bacchetta:2004it}
A.~Bacchetta and M.~Radici, {\it Dihadron interference fragmentation functions
  in proton- proton collisions},  {\em Phys. Rev.} {\bf D70} (2004) 094032,
  [\href{http://xxx.lanl.gov/abs/hep-ph/0409174}{{\tt hep-ph/0409174}}].

\bibitem{Yang:2009zzr}
{\bf PHENIX} Collaboration, R.~Yang, {\it {Transverse proton spin structure at
  PHENIX}},  {\em AIP Conf.Proc.} {\bf 1182} (2009) 569--572.

\end{thebibliography}\endgroup

\end{document}